\documentclass[seceq]{ptptex}
\usepackage{wrapft}
\usepackage{graphicx}

\newcommand{\vect}[1]{\overrightarrow{#1}}

\def\beq{\begin{eqnarray}}
\def\eeq{\end{eqnarray}}
\def\bsub{\begin{subequations}}
\def\esub{\end{subequations}}
\def\b{\begin{equation}}
\def\bs{\begin{split}}
\def\es{\end{split}}
\def\e{\end{equation}}
\begin{document}

\title{Spin Polarized versus Chiral Condensate  
in Quark Matter\\
at Finite Temperature and Density}

\author{Hiroaki {\sc Matsuoka}$^{1}$, Yasuhiko {\sc Tsue}$^{2}$, {Jo\~ao da {\sc Provid\^encia}}$^{3}$,\\ {Constan\c{c}a {\sc Provid\^encia}}$^{3}$,
{Masatoshi {\sc Yamamura}}$^{4}$ and {Henrik {\sc Bohr}}$^{5}$
%Thanks{These authors contributed equally to this work}}
}
%%%%%%%%%%% The \name command should be used as \name{Insert author name here}{Insert affiliation number here}
%%%%% Please use \thanks for contributed author details

%%%%%%%%%%% The \affil command should be used as \affil{Insert affiliation number here}{Insert author address here}
\inst{
$^{1}${Graduate School of Integrated Arts and Science, Kochi University, Kochi 780-8520, Japan}\\
$^{2}${Physics Division, Faculty of Science, Kochi University, Kochi 780-8520, Japan}\\
%\affil{2}{Departamento de F\'{i}sica, Universidade de Coimbra, 3004-516 Coimbra, 
%Portugal}\\
$^{3}${CFisUC, Departamento de F\'{i}sica, Universidade de Coimbra, 3004-516 Coimbra, 
Portugal}\\
$^{4}${Department of Pure and Applied Physics, 
Faculty of Engineering Science, Kansai University, Suita 564-8680, Japan}\\
$^{5}${Department of Physics, B.307, Danish Technical University, DK-2800 Lyngby, Denmark}
}

\abst{
It is shown that the spin polarized condensate appears in quark matter at high baryon density and low temperature 
due to the tensor-type four-point interaction in the Nambu-Jona-Lasinio-type model as a low energy effective theory of quantum chromodynamics. 
It is indicated within this low energy effective model that the chiral symmetry is broken again by the 
spin polarized condensate as increasing the quark number density, while the chiral symmetry restoration occurs in which the chiral condensate disappears 
at a certain density. 
}

%\subjectindex{xxxx, xxx}

\maketitle

\section{Introduction}

One of recent interests in the physics of the strong interaction, namely, in the physics governed by 
quantum chromodynamics (QCD), may be to clarify the structure of the phase diagram on the plane 
with respect to baryon chemical potential and temperature \cite{FH}. 
In the region of finite temperature and zero baryon chemical potential, lattice QCD simulation 
works and reliable calculations based on the first principles are performed until now. 
However, in the region of the low temperature and the finite baryon chemical potential, 
the possibility for various phases has been indicated such as the color superconducting phase \cite{CS,ARW,IB}, 
quarkyionic phase \cite{MaC}, inhomogeneous chiral condensed phase \cite{NT,Nic,Bub} and so on. 

%%%%%
%{\bf
In heavy-ion collision experiments such as the relativistic heavy-ion collider (RHIC) experiment, 
it is believed that quark-gluon phase is realized. 
Also, in the large hadron collider (LHC) experiment, it is expected that more extreme states of QCD 
with finite temperature and density and/or a strong magnetic field may be created in quark-gluon phase. 
It is interesting to understand what phases arise under extreme conditions. 
The quark-gluon phase under extreme conditions may be realized in the inner core of compact stars 
such as neutron stars, magnetars and quark stars if they exist. 
Therefore, the investigation of quark matter at low temperature and high density is also important to 
understand the compact star objects. 
%}
%%%%%%%%%%%%%

%
%{\bf 
In our previous papers,
%}
%On the other hand, 
it has been shown that 
a spin polarized phase may appear and be realized instead of the color superconducting phase 
in both the cases of two- \cite{oursPTEP1} or three-flavor  \cite{oursPTEP2}
in the region with finite quark chemical potential 
at zero temperature.
It is further interesting to investigate possible phases in the region with high density and low temperature 
from a viewpoint of physics of compact stars, especially, the structure of inner core of the compact stars. 
It has also been shown in our recent work \cite{oursPTEP3} that there is a possibility of the existence of  
a strong magnetic field on the surface of compact stars if there exists a quark spin polarized phase, 
which leads to the spontaneous magnetization of quark matter due to anomalous magnetic moment of quark, 
while only symmetric quark matter has been considered. 
If the spin polarization really leads to the spontaneous magnetization in the mechanism developed in the 
previous paper \cite{oursPTEP3}, it is a possible candidate for the origin of the strong magnetic field in 
so-called magnetar. \cite{magnetar1,magnetar2,magnetar3}

In this paper succeeding to Refs.\citen{oursPTP} and \citen{oursPTEP1}, 
a possibility of the quark spin polarized phase is investigated in the region of finite quark chemical potential 
and finite temperature by using the Nambu-Lona-Lasinio (NJL) model \cite{NJL,HK,Buballa} 
with the tensor-type four-point interaction between quarks \cite{arXiv}, instead of the 
pseudovector-type four-point interaction \cite{NMT,TMN}. 
As for the tensor-type four-point interaction, this interaction term was also introduced to investigate the meson spectroscopy, 
especially, for vector and axial-vector mesons \cite{tensor}. 
As another application, the dynamic properties of vector mesons were investigated in the extended NJL model including 
the tensor-type interaction \cite{tensor2}. 
Also, the chiral condensate and the quark spin polarization, namely, the tensor condensate, are considered simultaneously 
in the case with only one flavor 
instead of the color superconductor \cite{Ferrer}.

This paper is organized as follows: 
In the next section, the recapitulation of the NJL model with the tensor-type four-point interaction between quarks 
is given and in this model the chiral condensate and quark spin polarized condensate are considered simultaneously.  
In \S 3, the thermodynamic potential at zero temperature is     
introduced under the mean field approximation. 
In \S 4, the thermodynamic potential at finite temperature and density is given and derived. 
A derivation of the thermodynamic potential at zero temperature from that at finite temperature is given in Appendix A. 
Also, the effective potential is evaluated in Appendix B. 
%
%{\bf
In Appendix C, the analytic calculation for the thermodynamic potential is presented. 
%}
%
In \S 5, the numerical results are given through the calculation of the thermodynamic potential under 
various temperatures and quark chemical potentials.
The results are summarized in the phase diagram on the plane with respect to the quark chemical potential and temperature, 
in which the possible phases, the position of phase boundary  and the order of the phase transition are shown, apart from the 
color superconducting phase. 
%
%{\bf
In Appendix D, an idea introducing the tensor-type four-point interaction between quarks, which plays an essential role in this paper, 
is given from a viewpoint of two-gluon exchange process in QCD.
%}
%
The last section is devoted to a summary and concluding remarks.

\setcounter{equation}{0}

\section{NJL model with a tensor-type four-point interaction}
Let us consider the NJL-model Lagrangian density with a tensor-type four-point interaction.
The Lagrangian density with $su(2)$-flavor symmetry can be expressed as
\begin{align}
&\mathcal{L} = \mathcal{L}_0 + \mathcal{L}_{\text{S}} + \mathcal{L}_{\text{T}}, \\ 
&\mathcal{L}_0 = \bar{\psi} (i \gamma^\mu \partial_\mu - m_0)\psi, \\ 
&\mathcal{L}_{\text{S}} = G_\text{S} \left\{ (\bar{\psi} \psi)^2 + (\bar{\psi} i \gamma^5 \vec{\tau} \psi)^2 \right\}, \\
&\mathcal{L}_{\text{T}} = - \frac{G_\text{T}}{4} \left\{ (\bar{\psi} \gamma^\mu \gamma^\nu \vec{\tau} \psi) \cdot 
(\bar{\psi} \gamma_\mu \gamma_\nu \vec{\tau} \psi) + (\bar{\psi} i \gamma^5 \gamma^\mu \gamma^\nu \psi) 
(\bar{\psi} i \gamma^5 \gamma_\mu \gamma_\nu \psi) \right\},
\end{align}
where $m_0$ is a current quark mass for up and down quark and the components of $\vec{\tau}$ are the Pauli matrices for isospin.
It is known that these current quark masses are slightly different  for each flavor, but we have used the same value approximately.
The first two terms, $\mathcal{L}_0 + \mathcal{L}_{\text{S}}$, is the original NJL-model Lagrangian density.
In this paper $\mathcal{L}_{\text{T}}$ is added into the model, according to the Fierz transform.
Then, the spin matrix appears from $\mathcal{L}_{\text{T}}$ when $\mu = 1$, $\nu = 2$ or $\mu = 2$, $\nu = 1$ as follows:
\begin{equation*}
\varSigma_3 = -i \gamma^1 \gamma^2 = 
\begin{pmatrix} \sigma_3 & 0 \\ 0 & \sigma_3 \end{pmatrix} .
\end{equation*}
Since we use the mean field approximation, then we get the mean field Lagrangian density as
\begin{align}
\mathcal{L}_{\text{MFA}} = &\bar{\psi} (i \gamma^\mu \partial_\mu - m_0) \psi + 
G_{\text{S}} \left\{ 2 \langle \bar{\psi} \psi \rangle (\bar{\psi} \psi) - \langle \bar{\psi} \psi \rangle ^2 \right\} \notag \\
&+ \frac{G_{\text{T}}}{2} \left\{ 2 \langle \bar{\psi} \varSigma_3 \tau_3 \psi \rangle (\bar{\psi} \varSigma_3 \tau_3 \psi) 
- \langle \bar{\psi} \varSigma_3 \tau_3 \psi \rangle ^2 \right\},
\label{eq:lagrangian1}
\end{align}
where $\langle \cdots \rangle$ means vacuum expectation value.
Here $\tau_3$ is the third component of the Pauli matrix for isospin. 
When it operates on $\psi$ for up-quark (down-quark), 
the matrix changes into $1$ $(-1)$ as its eigenvalue.
Thus we could safely express $\bar{\psi} \varSigma_3 \tau_3 \psi$ as follows:
\[
\bar{\psi} \varSigma_3 \tau_3 \psi \rightarrow \bar{\psi} \varSigma_3 \psi \tau_f ,
\]
where $\tau_f = 1$ $(-1)$ when $f = $ up-quark (down-quark).
Let us define the following quantities:
\[
F \equiv -G_{\text{T}} \langle \bar{\psi} \varSigma_3 \psi \rangle, \, \qquad
M \equiv -2 G_{\text{S}} \langle \bar{\psi} \psi \rangle, \, \qquad
M_{\text{q}} \equiv m_0 + M.
\]
$F$ and $M$ are especially important quantities, because if $F$ and/or $M$ are not equal to zero, then spin polarization and/or chiral condensation occur.
Here, $M_\text{q}$ is just a constituent quark mass.
Substituting these quantities into Eq.(\ref{eq:lagrangian1}), we convert $\mathcal{L}_{\text{MFA}}$ into
\begin{equation}
\mathcal{L}_{\text{MFA}} = \bar{\psi} (i \gamma^\mu \partial_\mu - M_{\text{q}}) \psi 
- F (\bar{\psi} \varSigma_3 \psi) - \frac{M^2}{4 G_{\text{S}}} - \frac{F^2}{2 G_{\text{T}}} .
\end{equation}

Let us switch from Lagrangian formalism to Hamiltonian formalism by Legendre transformation.
First we must obtain the canonical momentum $\pi_\alpha$, 
$
\pi_\alpha = {\partial \mathcal{L}_{\text{MFA}}}/{\partial \dot{\psi_\alpha}} = i \psi_\alpha^\dagger
$,
where $\alpha$ means an index for spinor and isospin.
We, therefore, get the Hamiltonian density:
\begin{align}
\mathcal{H}_{\text{MFA}} &= \pi_\alpha \dot{\psi_\alpha} - \mathcal{L}_{\text{MFA}} \notag \\ 
&= \bar{\psi} (-i \vec{\gamma} \cdot \vec{\nabla} + M_{\text{q}}) \psi 
+ F (\bar{\psi} \varSigma_3 \psi) + \frac{M^2}{4 G_{\text{S}}} + \frac{F^2}{2 G_{\text{T}}} .
\end{align}
Thus, the Hamiltonian is expressed as
\begin{equation*}
H_{\text{MFA}} = \int d^3 x \, \psi^\dagger \gamma^0 (-i \vec{\gamma} \cdot \vec{\nabla} + M_{\text{q}} + F \varSigma_3) \psi 
+ V \frac{M^2}{4 G_{\text{S}}} + V \frac{F^2}{2 G_{\text{T}}} ,
\end{equation*}
where V is the volume of this system.
We transform $\psi(x)$ by the Fourier transformation as 
$\psi(x) = \int {d^3 p}/{(2 \pi)^3} \cdot\tilde{\psi}(p) e^{i {\vec p} \cdot {\vec x}}.
$
Substituting it into $H_{\text{MFA}}$, we obtain
\begin{equation}
H_{\text{MFA}} = \int \frac{d^3 p}{(2 \pi)^3} \tilde{\psi}^\dagger \gamma^0 (\vec{\gamma} \cdot \vec{p} + M_{\text{q}} + F \varSigma_3) \tilde{\psi}
+ V \frac{M^2}{4 G_{\text{S}}} + V \frac{F^2}{2 G_{\text{T}}} .
\end{equation}
What we must do is to diagonalize $H_{\text{MFA}}$.
Non diagonal terms are
\begin{align*}
h_{\text{MFA}} &\equiv \gamma^0 (\vec{\gamma} \cdot \vec{p} + M_{\text{q}} + F \varSigma_3) \notag \\
&= \begin{pmatrix} F \sigma_3 + M_{\text{q}} & \vec{p} \cdot \vec{\sigma} \\ \vec{p} \cdot \vec{\sigma} & -F \sigma_3 - M_{\text{q}} \end{pmatrix} \notag \\
&= \begin{pmatrix} 
F + M_{\text{q}} & 0 & p_3 & p_1 - i p_2 \\ 0 & -F + M_{\text{q}} & p_1 + i p_2 & -p_3 \\
p_3 & p_1 - i p_2 & -F - M_{\text{q}} & 0 \\ p_1 + i p_2 & -p_3 & 0 & F - M_{\text{q}}
\end{pmatrix}.
\end{align*}
Since $h_{\text{MFA}}$ is a hermitian matrix, it is diagonalized by a unitary matrix.
The eigenvalues are obtained as
\begin{equation}
\pm E^{(\eta)}_{\vec{p}} = \pm \sqrt{p_3^2 +\left(\sqrt{p_1^2 + p_2^2 + M_{\text{q}}^2} + \eta F \right)^2},
\end{equation}
where $\eta = \pm 1$.

\setcounter{equation}{0}

\section{Thermodynamic potential at zero temperature}

Next, we introduce a quark chemical potential $\mu$ and a number density operator $\mathcal{N}$ in order to discuss on a finite density system at 
zero temperature.
The thermodynamic potential is defined as follows:
\begin{equation}
\varPhi = \mathcal{H}_{\text{MFA}} - \mu \mathcal{N}.
\end{equation}
The next step is to calculate the expectation value.
Since we consider the zero-temperature system in this section, the system that quasi-particles degenerate is treated.
Hence, we must sum over momenta from zero to the single-quasi-particle energy equal to the chemical potential.
Sandwiching with a ``bra'' and ``ket'', we obtain
\begin{align*}
\varPhi&=\frac{1}{V} \langle \text{F.D.} | \left( H_{\text{MFA}} -\mu \int d^3 x \, \mathcal{N} \right) | \text{F.D.} \rangle \\ 
&=\frac{1}{V} \sum^{E^{(\eta)}_{\vec{p}} \le \mu, \vec{p}^2 \le \varLambda^2}_{\vec{p}, \eta, \tau, \alpha} \left( E^{(\eta)}_{\vec{p}} - \mu \right) 
+ \frac{M^2}{4 G_{\text{S}}} + \frac{F^2}{2 G_{\text{T}}},
\end{align*}
where $| \text{F.D.} \rangle$ means the degenerate Fermi gas constituted by quasi-particles and $\varLambda$ is a three-momentum cutoff parameter 
for the integration over momenta.
Here, $\tau$ and $\alpha$ are indeces for isospin and quark color, respectively.
The upper limit of integration is imposed by two conditions, which are $E^{(\eta)}_{\vec{p}} \le \mu$ and $\vec{p}^2 \le \varLambda^2$.
We would like to discuss spin polarization and chiral condensate simultaneously. 
However, the above expression does not have the contribution from Dirac sea.
Since chiral condensate occurs by the effect of Dirac sea, we must add its contribution.
Thus, we get
\begin{equation}
\varPhi(M, F, \mu)=\frac{1}{V} \sum^{E^{(\eta)}_{\vec{p}} \le \mu, \vec{p}^2 \le \varLambda^2}_{\vec{p},\eta,\tau,\alpha} \left( E^{(\eta)}_{\vec{p}}-\mu \right) 
- \frac{1}{V} \sum^{\vec{p}^2 \le \varLambda^2}_{\vec{p}, \eta, \tau, \alpha} E^{(\eta)}_{\vec{p}} + \frac{M^2}{4 G_{\text{S}}} + \frac{F^2}{2 G_{\text{T}}},
\end{equation}
where, the second term represents the contribution from the Dirac sea (negative energy sea).
We change the sum $\frac{1}{V} \sum_{\vec{p}}$ into the integration $\int \frac{d^3 p}{(2 \pi)^3}$. 
Then, the thermodynamic potential can be expressed as
\begin{equation}
\varPhi(M, F, \mu) = \varPhi_1 + \varPhi_2 + \varPhi_3 + \varPhi_4,
\label{eq:thermodynamic1}
\end{equation}
where,
\begin{align}
\varPhi_1(F, M, \mu) = \sum_{\tau, \alpha} &\int_{\varGamma_1} \frac{d^3 p}{(2 \pi)^3} \left \{ 
\sqrt{p_3^2+(\sqrt{p_1^2 + p_2^2 + M_{\text{q}}^2} + F)^2} - \mu \right \}, \notag \\
&\varGamma_1 = \left\{ E^{(+1)}_{\vec{p}} \le \mu, \, \vec{p}^2 \le \varLambda^2 \right \} 
\end{align}
\begin{align}
\varPhi_2(M, F, \mu) = \sum_{\tau, \alpha} &\int_{\varGamma_2} \frac{d^3 p}{(2 \pi)^3} \left \{ 
\sqrt{p_3^2+(\sqrt{p_1^2 + p_2^2 + M_{\text{q}}^2} - F)^2} - \mu \right \}, \notag \\
&\varGamma_2 = \left\{ E^{(-1)}_{\vec{p}} \le \mu, \, \vec{p}^2 \le \varLambda^2 \right \} 
\end{align}
\begin{align}
\varPhi_3(M, F, \mu) = -\sum_{\eta, \tau, \alpha} &\int_{\varGamma_3} \frac{d^3 p}{(2 \pi)^3} 
\sqrt{p_3^2+(\sqrt{p_1^2 + p_2^2 + M_{\text{q}}^2} + \eta F)^2}, \ \ \ \notag \\ 
&\varGamma_3 = \left\{ \vec{p}^2 \le \varLambda^2 \right \} \quad
\end{align}
\begin{equation}
\varPhi_4(M,F,\mu) = \frac{M^2}{4 G_{\text{S}}} + \frac{F^2}{2 G_{\text{T}}}. \qquad\qquad\qquad\qquad\qquad\qquad\qquad\qquad
\end{equation}
Here, $\varPhi_i$ $(i = 1, 2, 3, 4)$ means, respectively, the contribution from positive energy for $\eta = +1$, from positive energy for $\eta = -1$,
from vacuum and the mean field contribution, respectively.
$\varGamma_i$ $(i = 1, 2, 3)$ is the domain of integration over momenta.
Since these integrands do not depend on $\tau$ or $\alpha$, the summations over $\tau$ and $\alpha$ give factors $2$ and $3$, respectively.

\setcounter{equation}{0}

\section{Thermodynamic potential at finite temperature and density}

We have discussed the thermodynamic potential at zero temperature in the previous section.
In this section let us consider the thermodynamic potential at finite temperature.
We define a thermodynamic potential at finite temperature, $\varOmega(M,F,\mu,T)$, as follows:
\begin{align}
&\varOmega(M,F,\mu,T) \equiv \mathcal{H}'_{\text{MFA}}-\mu(\mathcal{N}_{\text{P}}-\mathcal{N}_{\text{AP}})+V_{\text{vacuum}}-TS,\\
&\mathcal{H}'_{\text{MFA}} \equiv \sum_{\vec{p},\eta,\tau,\alpha} E^{(\eta)}_{\vec{p}} \left( n^{(\eta)}_{\vec{p}}+\bar{n}^{(\eta)}_{\vec{p}} \right)
+\frac{M^2}{4G_{\text{S}}}+\frac{F^2}{2G_{\text{T}}}, \notag \\
&\mathcal{N}_{\text{P}} \equiv \sum_{\vec{p},\eta,\tau,\alpha} n^{(\eta)}_{\vec{p}},\, \qquad
\mathcal{N}_{\text{AP}} \equiv \sum_{\vec{p},\eta,\tau,\alpha} \bar{n}^{(\eta)}_{\vec{p}}, \notag \\
&n^{(\eta)}_{\vec{p}}=\frac{1}{1+\exp \left( (E^{(\eta)}_{\vec{p}}-\mu)/T \right)},\, \qquad
\bar{n}^{(\eta)}_{\vec{p}}=\frac{1}{1+\exp \left( (E^{(\eta)}_{\vec{p}}+\mu)/T \right)}, \notag \\
&V_{\text{vacuum}} \equiv -\sum_{\vec{p},\eta,\tau,\alpha} E^{(\eta)}_{\vec{p}},\notag
\end{align}
where $T$ means temperature of the system, $n^{(\eta)}_{\vec{p}}$ and $\bar{n}^{(\eta)}_{\vec{p}}$ are the distribution functions for particle and anti-particle,
respectively.
The entropy $S$ is given as follows:
\begin{align*}
S&=-\sum_{\vec{p},\eta,\tau,\alpha} \biggl\{ n^{(\eta)}_{\vec{p}} \log n^{(\eta)}_{\vec{p}}+(1-n^{(\eta)}_{\vec{p}})\log(1-n^{(\eta)}_{\vec{p}}) \\
&\qquad\qquad\quad
+\bar{n}^{(\eta)}_{\vec{p}} \log \bar{n}^{(\eta)}_{\vec{p}}+(1-\bar{n}^{(\eta)}_{\vec{p}})\log(1-\bar{n}^{(\eta)}_{\vec{p}}) \biggr\}.
\end{align*}
Using the following identities
\begin{align*}
&n^{(\eta)}_{\vec{p}} \log n^{(\eta)}_{\vec{p}}=-n^{(\eta)}_{\vec{p}} \times \frac{E^{(\eta)}_{\vec{p}}-\mu}{T}
-n^{(\eta)}_{\vec{p}}\log(1-n^{(\eta)}_{\vec{p}}),\\
&\bar{n}^{(\eta)}_{\vec{p}} \log \bar{n}^{(\eta)}_{\vec{p}}=-\bar{n}^{(\eta)}_{\vec{p}} \times \frac{E^{(\eta)}_{\vec{p}}+\mu}{T}
-\bar{n}^{(\eta)}_{\vec{p}}\log(1-\bar{n}^{(\eta)}_{\vec{p}}),
\end{align*}
the above $\varOmega$ can be recast into
\begin{align*}
\varOmega&=-\sum_{\vec{p},\eta,\tau,\alpha} \left\{ E^{(\eta)}_{\vec{p}}
+T \log \left( 1+\exp \left( -\frac{E^{(\eta)}_{\vec{p}}-\mu}{T} \right) \right)
+T \log \left( 1+\exp \left( -\frac{E^{(\eta)}_{\vec{p}}+\mu}{T} \right) \right)\right\} \\
&\ \ +\frac{M^2}{4G_{\text{S}}}+\frac{F^2}{2G_{\text{T}}}.
\end{align*}
Let us change the summation over momenta into integration, and then let us introduce polar coordinates,
$p_1 = p_T \cos \theta,\,p_2 = p_T \sin \theta$, so as to integrate.
The domain of integration is obtained as follows:
\begin{equation*}
-\sqrt{\varLambda^2-p_T^2} \le p_3 \le \sqrt{\varLambda^2-p_T^2},\,\qquad
0 \le p_T \le \varLambda.
\end{equation*}
After integrating over $\theta$, and summing over $\tau$ and $\alpha$, we get the final form:
\begin{align}\label{5-2}
\varOmega(M,F,\mu,T)&=-\frac{3}{\pi^2} \sum_{\eta} \int_{0}^{\varLambda} dp_T \int_{0}^{\sqrt{\varLambda^2-p_T^2}}dp_3\,p_T \notag \\
&\qquad\qquad
\times \left\{ E^{(\eta)}_{\vec{p}}+T \log \left( 1+\exp \left( -\frac{E^{(\eta)}_{\vec{p}}-\mu}{T} \right) \right)
\right.
\nonumber\\
&\qquad\qquad\qquad\left.
+T \log \left( 1+\exp \left( -\frac{E^{(\eta)}_{\vec{p}}+\mu}{T} \right) \right) \right\} \notag \\
&+\frac{M^2}{4G_{\text{S}}}+\frac{F^2}{2G_{\text{T}}}.
\end{align}

%%%%%%%%%%%%%%%%%%%%%%%%%%%%%%
\begin{table}[b]
\caption{Parameter set }
\label{table:1}
\begin{center}
\begin{tabular}{c|c|c|c} \hline
{
$\varLambda$/\text{GeV}} & { $m_0$/\text{GeV}} & { $G_{\text{S}}$/\text{GeV}$^{-2}$} & { $G_{\text{T}}$/\text{GeV}$^{-2}$}\\ \hline\hline
0.631 & 0.0 & 5.5 & 11.0 \\
\hline
\end{tabular}
\end{center}
\end{table}
%%%%%%%%%%%%%%%%%%%%%%%%%%%

\setcounter{equation}{0}

\section{Numerical results and discussions}

In this section we would like to discuss the thermodynamic potential at zero/finite temperature numerically.
In order to evaluate it we use the three-momentum cutoff parameter and coupling constants in Table 1.
%\begin{table}
%\caption{Parameter set.}
%\begin{center}
%	\begin{equation*}
%	\begin{array}{cccc}
%	\toprule
%	\varLambda/\text{GeV} & m_0/\text{GeV} & G_{\text{S}}/\text{GeV}^{-2} & G_{\text{T}}/\text{GeV}^{-2} \\
%	\midrule
%	0.631 & 0.0 & 5.5 & 11.0 \\
%	\bottomrule
%	\end{array}
%	\end{equation*}
%\end{center}
%\end{table}
Here, we adopt the strength of tensor interaction $G_{\rm T}$ as a rather small value compared with the one used in our previous paper. 
The reason why we take $G_{\rm T}$ as a rather small value 11.0 GeV is that the vacuum polarization, namely the contribution of the negative energy sea, is 
taken into account. 
A detailed discussion of this effect has already been given in appendix B in \cite{oursPTP}.
%
%{\bf
Here, in this section, we show numerical results in the case of the chiral limit, $m_0=0.0$.
%}

\subsection{Thermodynamic potential at zero temperature}

Let us discuss the thermodynamic potential at zero temperature.
First we consider the chiral condensate $M$ and the spin polarization $F$ separately.

\begin{figure}[t]
\centering
\includegraphics[height=5cm]{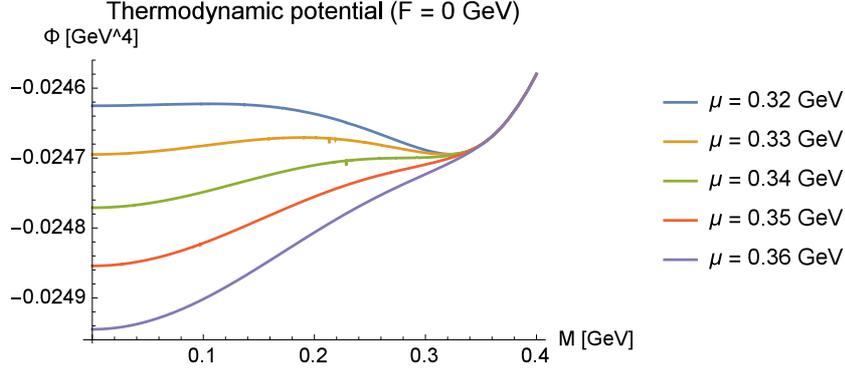}
\caption{The thermodynamic potential with $F=0$ is depicted as a function of the 
constituent quark mass $M$ in various quark chemical potentials. 
%horizontal and vertical axes represent $M$ and $\varPhi(M,\mu)$, respectively.
%The chemical potential is changed from $0.32$ GeV to $0.36$ GeV
}
\label{fig:fig1}
\end{figure}

Figure \ref{fig:fig1} shows the thermodynamic potential in the special case where $F=0$.
When the chemical potential has a value below $0.32$ GeV and above $0.35$ GeV, the thermodynamic 
potential has only one minimum.
On the other hand, when $\mu=0.33 \sim 0.34$ GeV, the thermodynamic potential has two local minima.
This indicates that the phase transition to chiral condensate is of first order.

%%%%%%%%%%%%%%%%%%%%%%%%%%%%%%%%%%%%%%%%%%%%%%
%%%%%%%%%%%%$$$$$$$$$$$$$$$$$$$$$$$$$$$$$$$$$$$$$$$$$$$$$$

Figure {\ref{fig:fig2} shows the thermodynamic potential for $M=0$.
When the chemical potential is small, the spin polarized phase does not appear. 
However, the chemical potential $\mu$ has a value above $0.42$ GeV, the spin polarized phase appears.
This figure shows that the phase transition to spin polarization is of the second order.

Next, let us consider $M$ and $F$ simultaneously.
In Fig.{\ref{fig:fig3}, the contour map for the thermodynamic potential 
is depicted with various quark chemical potentials. 
%and change the value of the chemical potential.
The horizontal and the vertical axes represent the constituent quark mass $M$ and the spin polarized condensate $F$, respectively. 
When $\mu$ varies from $0$ GeV to $0.32$ GeV, the chiral condensed phase arises. 
However, when $\mu$ reaches $0.35$ GeV, chiral symmetry is restored.
If $\mu=0.43$ GeV, the spin polarized phase appears.
From these contour maps, it indicates that two phases, the chiral condensed and the spin polarized phases, do not coexist.

%%%%%%%%%%%%%%%%%%%%%%%%%%%%%%%%%%%%%%%%%%%%%
\begin{figure}[t]
\centering
\includegraphics[height=5cm]{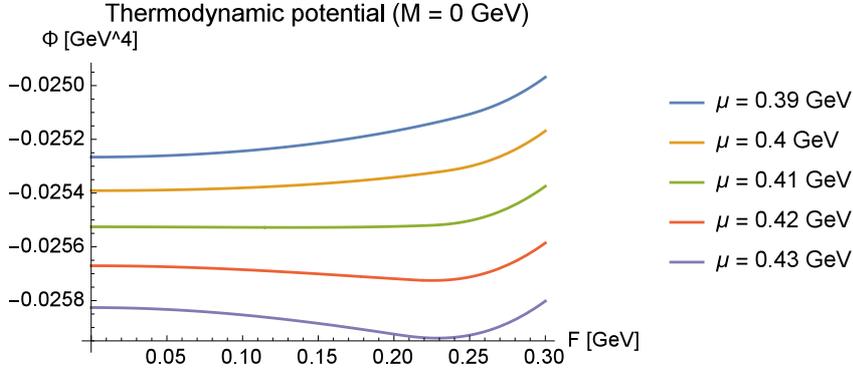}
\caption{
The thermodynamic potential with $M=0$ is depicted as a function of the 
spin polarized condensate $F$ in various quark chemical potentials. 
%The horizontal and vertical axes represent $F$ and $\varPhi(F,\mu)$, respectively.
%The chemical potential is changed from $0.39$ GeV to $0.43$ GeV
}
\label{fig:fig2}
\end{figure}
%%%%%%%%%%%%%%%%%%%%%%%%%%%%%%%%%%%%%%%%%%%%%

%%%%%%%%%%%%%%%%%%%%%%%%%%%%%%%%%%%%%%%%%%%%%%%%%%%%%%
\begin{figure}[t]
\centering
\includegraphics[height=10cm]{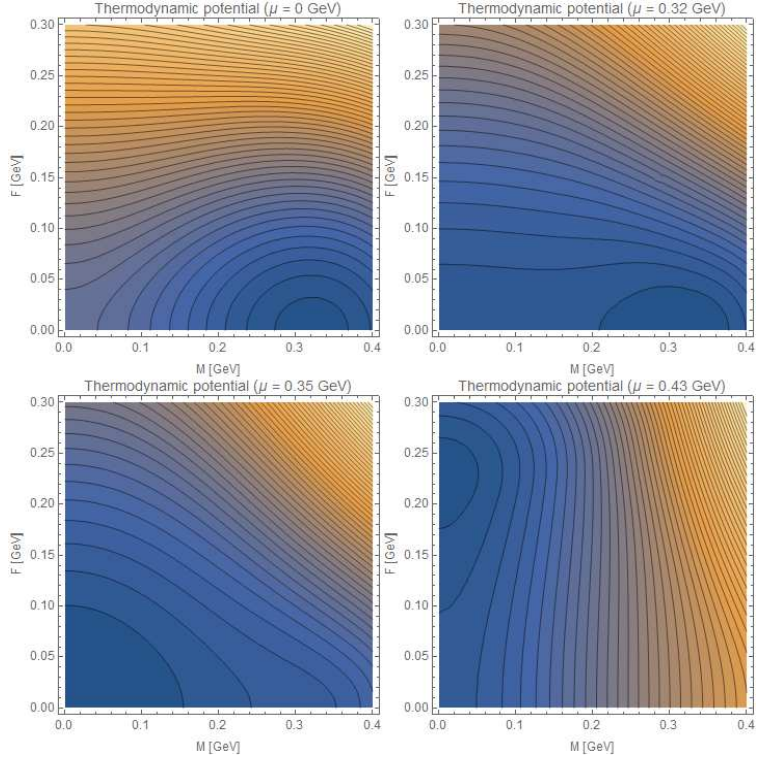}
\caption{
The contour map of the thermodynamic potential is depicted as a function of the 
constituent quark mass $M$ and the spin polarized condensate $F$ in various quark chemical potentials. 
The horizontal and vertical axes represent $M$ and $F$, respectively.
%We change the chemical potential from 0 GeV to 0.43 GeV.
As color becomes dark, the thermodynamic potential lowers.}
\label{fig:fig3}
\end{figure}
%%%%%%%%%%%%%%%%%%%%%%%%%%%%%%%%%%%%%%%%%%%%%%%%%%

%%%%%%%%%%%%%%%%%%%%%%%%%%%%%%%%%%%%%%%%%%%%%%%%%%%
\begin{figure}[t]
\centering
\includegraphics[width=\columnwidth]{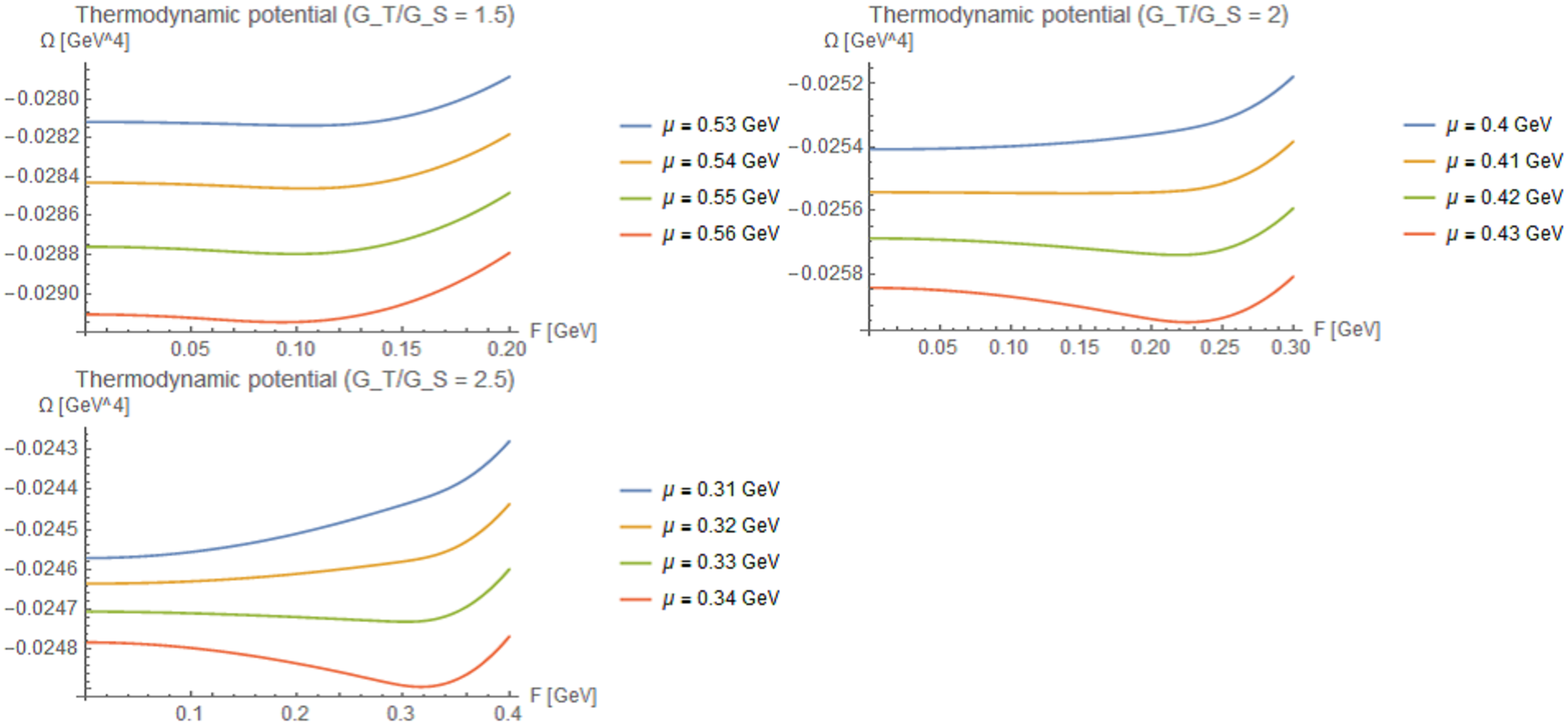}
\caption{The thermodynamic potentials with $M=0$ and $T=0.01$ are depicted in the various ratios of $G_T / G_S$.
The ratios $G_T / G_S = 1.5,\, 2$ and $2.5$ mean $G_T = 8.25,\, 11.0$ and $3.75 \ \text{GeV}^{-2}$, respectively.
}
\label{fig:fig4}
\end{figure}
%%%%%%%%%%%%%%%%%%%%%%%%%%%%%%%%%%%%%%%%%%%%%%%%%

%
%{\bf
Here, two notes should be mentioned. 
One is about the effect of $G_{T}$, namely the coupling strength of the tensor-type interaction. 
As was already mentioned in the beginning of this section, 
a detailed discussion about $G_T$ has already been given in appendix B in \cite{oursPTP}.
However, let us demonstrate the effect of $G_T$ for the spin polarization. 
Figure \ref{fig:fig4} shows the thermodynamic potential with $M=0$ in various values of $G_T$, namely,
$G_T/G_S=1,5,\ 2.0$ and $2.5$, respectively. 
As the coupling strength $G_T$ is increased, the critical chemical potential of phase transition is decreasing. 
For example, in the small value of $G_T$ such as $G_T/G_S=1.5$, the phase transition occurs around $\mu=0.54$ GeV. 
On the other hand, in the large $G_T$, $G_T/G_S=2.5$, the phase transition occurs around $\mu=0.33$ GeV.   
In this paper, we adopt a moderate value $G_T/G_S=2.0$ for discussions. 
%}
%%%%

Another is about a reason why the spin polarization occurs at large chemical potential. 
At zero temperature, the spin polarized phase is actually realized in our model. 
It is easy to understand how the spin polarized  phase appears. 
Neglecting the contribution of the chiral condensate, 
the energy $E$ of the system under consideration can be expressed by using the quark chemical potential as 
\beq\label{6-1}
E&=&\int^{\mu}\mu\frac{\partial N}{\partial \mu}d\mu 
+\frac{F^2}{2G_T}\nonumber\\
&=&\mu N-\int N d\mu +\frac{F^2}{2G_T}\ . 
\eeq
Thus, the thermodynamical potential $\Phi\ (=E-\mu N)$ is obtained as 
\beq\label{6-2}
\Phi&=&
E-\mu N\nonumber\\
&=&-\int N d\mu +\frac{F^2}{2G_T}
\nonumber\\
&=&-\int d\mu \sum_{\eta,\tau,\alpha}\int \frac{d^3{\mib p}}{(2\pi)^3}\theta(\mu-E_{\vec p}^{(\eta)})
 +\frac{F^2}{2G_T}\ ,  
\eeq
where $\theta(x)$ represent the Heaviside step function.
Let us consider the normal quark matter in which $F=0$. 
In this case, the above-derived thermodynamical potential can be calculated easily as 
\beq\label{6-3}
\Phi
%&=&
%-\frac{2}{\pi^2}\int^{\mu}d\mu
%\left[\frac{\mu^3}{3}+\frac{\Lambda^3}{3}+
%\int_0^{\mu}dp p^2-\int_{\mu}^{\Lambda} dp p^2\right]\nonumber\\
&=&-\frac{\mu^4}{2\pi^2}\ . 
\eeq
On the other hand, in the case $F\neq 0$, we can also calculate the thermodynamic potential analytically, which 
was presented in \cite{arXiv}. 
For simplicity, let us consider the case $F>\mu$. 
For $F>\mu$, $E_{\vec p}^{(+)}$ does not contributes to the three-momentum integration. 
In this case, by the existence of the theta function, the integration has a finite value in 
$E_{\vec p}^{(-)} \leq \mu$: 
\beq\label{6-4}
E_{\vec p}^{(-)}\equiv 
\sqrt{p_3^2+\left(F-\sqrt{p_1^2+p_2^2}\right)^2}\leq \mu. 
\eeq
In the case of equality in the above expression, this equality represents the 
formula of torus in which the major radius is $F$ and small radius is $\mu$. 
Thus, the Fermi surface has a form of torus. 
Therefore, the momentum integral $\int d^3{\bf p}\theta(\mu-E_{\vec p}^{(-)})$ 
means the volume of Fermi ``torus", where the volume gives 
$2\pi^2\mu^2 F \ (=\pi \mu^2\cdot 2\pi F)$. 
Then, we obtain the thermodynamical potential in the large $\mu$ region as 
\beq\label{6-5}
\Phi&=&
-\frac{3}{4\pi^3}\int^\mu d\mu 2\pi^2\mu^2 F+\frac{F^2}{2G_T}\nonumber\\
&=&-\frac{\mu^3 F}{2\pi}+\frac{F^2}{2G_T}\ .
\eeq
The ``gap equation" for $F$ is derived from 
\beq\label{6-6}
\frac{\partial \Phi}{\partial F}=-\frac{\mu^3}{2\pi}+\frac{F}{G_T}=0\ , \qquad
{\rm thus,}\qquad
F=\frac{G_T\mu^3}{2\pi}\ .
\eeq
Inserting the above-derived $F$ into the thermodynamical potential, 
we finally obtain 
\beq\label{6-7}
\Phi=-\frac{G_T\mu^6}{8\pi^2}\ . 
\eeq
For small chemical potential, namely, at low quark number density, 
normal quark matter is realized because the thermodynamic potential has order 
$\mu^4$. 
On the other hand, for large chemical potential, namely, at high quark number density, 
the thermodynamical potential with the order of $\mu^6$ overcomes the 
normal quark matter with the order of $\mu^4$. 
It may be concluded that the appearance of the spin polarized phase is due to the effect of the volume of the phase space. 
Thus, at high density, the spin polarized phase is realized absolutely. 
%%%%

%%%%%%%%%%%%%%%%%%%%%%%%%%%%%%%%%%%%%%%%%%%%%%%%%%%
\begin{figure}[b]
\centering
\includegraphics[height=7cm]{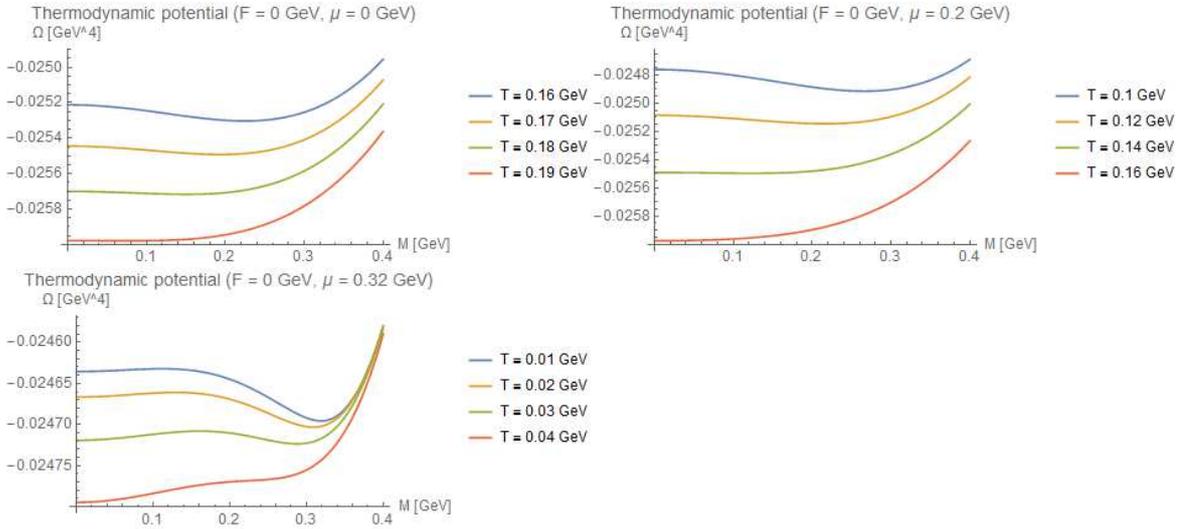}
\caption{
The thermodynamic potentials with $F=0$ are depicted in the various temperatures $T$ in the case with the chemical potential $\mu=0$, $0.2$ GeV and $0.32$ 
GeV, respectively.
The horizontal and vertical axes represent the constituent quark mass $M$ and the thermodynamic potential $\varOmega(M,\mu,T)$, respectively.
%We plot for $\mu=0$ GeV, $\mu=0.32$ GeV and $\mu=0.32$ GeV, and change temperature $T$.
}
\label{fig:fig5}
\end{figure}
%%%%%%%%%%%%%%%%%%%%%%%%%%%%%%%%%%%%%%%%%%%%%%%%%

%%%%%%%%%%%%%%%%%%%%%%%%%%%%%%%%%%%%%%%%%%%%%%%%%%%%%%%%
\begin{figure}[b]
\centering
\includegraphics[height=7cm]{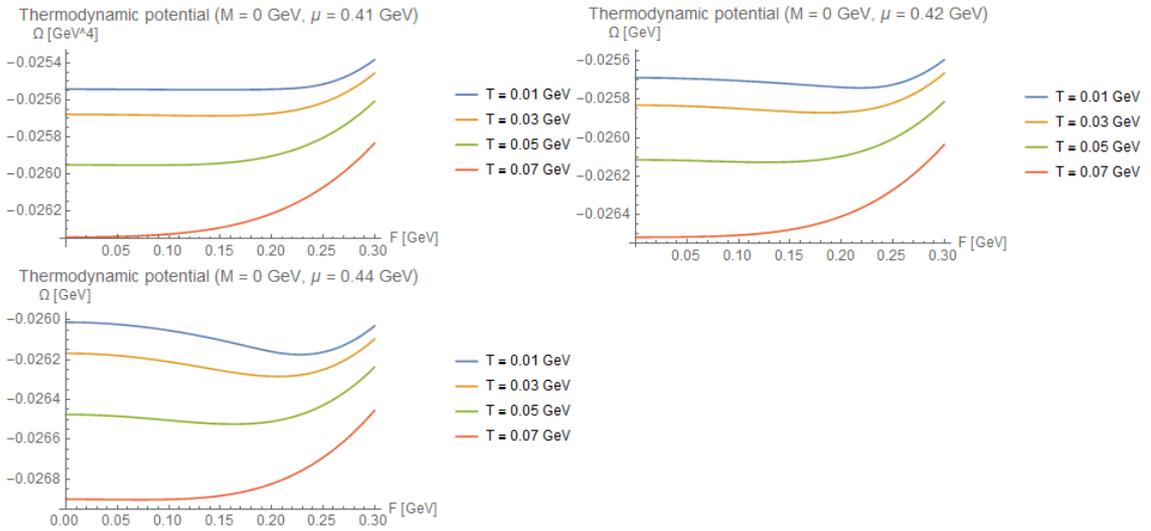}
\caption{
The thermodynamic potentials with $M=0$ are depicted in the various temperatures $T$ in the case with the chemical potential $\mu=0.41$ GeV, 
$0.42$ GeV and $0.44$ GeV, respectively.
The horizontal and vertical axes represent the spin polarized condensate $F$ and the thermodynamic potential $\varOmega(F,\mu,T)$, respectively.
%The horizontal and vertical axes represent $F$ and $\varPhi(F,\mu,T)$, respectively.
%We plot when $\mu=0.41$ GeV, $\mu=0.42$ GeV and $\mu=0.44$ GeV, and temperature is changed from $T=0.01$ GeV to $T=0.07$ GeV
.}
\label{fig:fig6}
\end{figure}
%%%%%%%%%%%%%%%%%%%%%%%%%%%%%%%%%%%%%%%%%%%%%%%%%%%%%%%%

%%%%%%%%%%%%%%%%%%%%%%%%%%%%%%%%%%%%%%%%%%%%%%%%%%%%%%%%%%%%%
\begin{figure}[t]
\centering
\includegraphics[height=13cm]{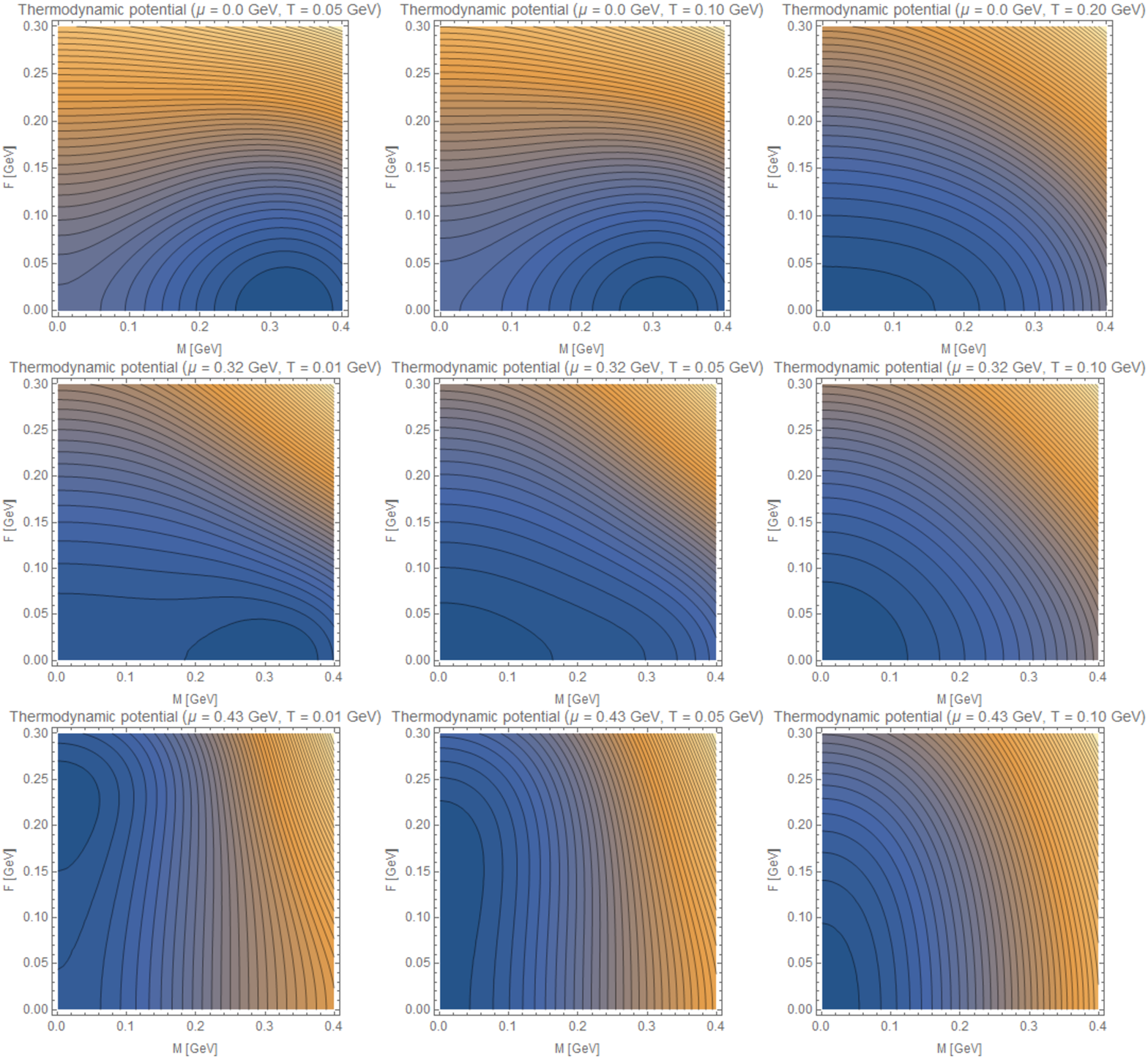}
\caption{
The contour maps of the thermodynamic potential are depicted as a function of 
the constituent quark mass $M$ and the spin polarized condensate $F$ with various quark chemical potentials and temperatures.  
The horizontal and vertical axes represent $M$ and $F$, respectively. 
%We plot when $\mu=0$ GeV, $\mu=0.32$ GeV and $\mu=0.43$ GeV.
%Temperature $T$ is also changed.
}
\label{fig:fig7}
\end{figure}
%%%%%%%%%%%%%%%%%%%%%%%%%%%%%%%%%%%%%%%%%%%%%%%%%%%%%%%%%%%%

\subsection{thermodynamic potential at finite temperature}

Let us consider the thermodynamic potential at finite temperature.
First, let us treat two cases without $M$ or $F$ separately.
Figure \ref{fig:fig5} shows the thermodynamic potential at finite temperature for $F=0$.
If temperature $T$ is not so high, the chiral condensed phase is realized. 
However, in high temperature region the chiral condensed phase  disappears.
It should be noted that, in the cases with $\mu=0$ GeV and $\mu=0.2$ GeV, the phase transition is of second order, 
while the phase transition is of first order in the case $\mu=0.32$ GeV. 

Secondly, we discuss the case for $M=0$.
In Fig. \ref{fig:fig6}, it is shown that the spin polarized phase is realized in only low temperature region.
If temperature rises, the spin polarization disappears soon.
In this case, the phase transition from the spin polarized phase to the normal phase is of second order.

Finally, let us consider $M$ and $F$ simultaneously. 
% with current quark mass.
It is shown in Fig. \ref{fig:fig7} that chiral symmetry is broken for small chemical potential and low temperature.
However, if the chemical potential or temperature becomes high, chiral symmetry is restored.
In the large chemical potential and low temperature region, the spin polarized condensate appears. 
However, for higher temperature, it disappears.
According to these contour maps, it may be concluded that the two phases, 
namely the chiral condensed phase and the spin polarized phase, do not coexist at finite temperature.

\subsection{Phase diagram on $T$-$\mu$ plane}

As a summary, it is possible to show the regions of the chiral condensed phase and the spin polarized phase 
on the plane with the temperature $T$ and the quark chemical potential $\mu$ and also to draw the phase boundary 
indicating the order of phase transition under the chiral limit, $m_0=0$.  
%We plot the point where chiral condensate phase or spin polarized phase disappears.
In Fig. \ref{fig:fig8}, the phase diagram in this model is presented.
As is shown in this phase diagram, the chiral condensed phase exists in the left side on the $T$-$\mu$ plane 
and the spin polarized phase exists in the right side. 
It is indicated that, for the boundary of the chiral condensed and the normal phases, there is a critical endpoint for the phase transition 
near $\mu=0.31$ GeV and $T=0.046$ GeV. 
%and it disappears at $\mu=0.33$ GeV and zero temperature.
%%%%
%{\bf 
On the other hand, the phase transition from the normal quark phase to the spin polarized phase is always of second order 
and there is no endpoint. 
%}
%%%%

%%%%%%%%%%%%%%%%%%%%%%%%%%%%%%%%%%%%%%%%%%%%%%%%%%%%%%%%%%%%%%%%%%%%%%%%%%%%
\begin{figure}[t]
\centering
\includegraphics[height=6.5cm]{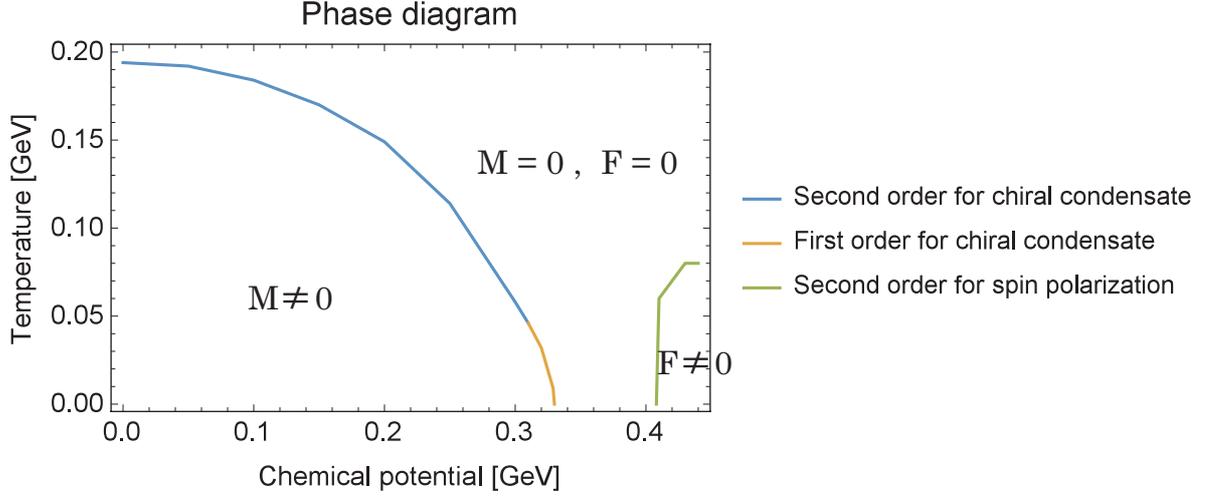}
\caption{
The phase diagram in this model is depicted. 
The horizontal and vertical axes represent the quark chemical potential and the temperature, respectively. 
%This is the phase diagram in $T$-$\mu$ plane.
}
\label{fig:fig8}
\end{figure}
%%%%%%%%%%%%%%%%%%%%%%%%%%%%%%%%%%%%%%%%%%%%%%%%%%%%%%%%%%%%%%%%%%%%%%%%%

\setcounter{equation}{0}

\section{Summary and concluding remarks}

In this paper, it has been shown that the spin polarized phase appears in the region with the large quark chemical potential and 
low temperature by using the NJL model with tensor-type four-point interaction between quarks. 
We have considered the chiral condensate and spin polarized condensate simultaneously. 
For rather low density, the chiral condensate exists and spin polarized condensate does not exist. 
As the quark chemical potential is increased, the chiral condensate disappears and further, the spin polarized condensate 
arises.
Thus, the spin polarized phase may exist in the high density and low temperature region in QCD phase diagram. 
However, it may be concluded that the two phases do not coexist in this model under the parameter set adopted here. 

It should be also indicated that the color superconducting phase may be realized in the region with high density and low temperature. 
However, at zero temperature, the spin polarized phase may be realized instead of the two-flavor color superconducting phase in the case with 
only two flavors \cite{oursPTEP1}. 
It is interesting that the spin polarized phase survives or not at finite temperature instead of the color superconducting phase. 
It is one of future important problems to investigate. 
%Thus, it is interesting to study the relation between spin polarization and color superconductivity.
Furthermore, in this paper we do not consider the electromagnetic field at all. 
It is also important to study the electromagnetic properties in the spin polarized phase, 
%%%%%%%%%%%%
%{\bf 
for example, the spontaneous magnetization in compact stars.
Especially, the charge neutrality and $\beta$-equilibrium play an essential role to discuss the 
physics of neutron stars and/or magnetars. 
It is also one of interesting future problems.
%}
%%%%%%%%%%%%%% 
%We are interested in the relation between tensor-type and pseudovector-type interaction, too.

\section*{Acknowledgment}

%One of the authors (Y.T.) would like to express his sincere thanks to 
%Professor\break
%J. da Provid\^encia and Professor C. Provid\^encia, two of co-authors of this paper, 
%for their warm hospitality during his visit to Coimbra in spring of 2014. 
%Two of the authors (J.P. and C.P.) acknowledge valuable
%discussions with H. Bohr and\break
%P. K. Panda.
One of the authors (Y.T.) 
is partially supported by the Grants-in-Aid of the Scientific Research 
(No.26400277) from the Ministry of Education, Culture, Sports, Science and 
Technology in Japan.

\appendix

\setcounter{equation}{0}

\section{Derivation of the thermodynamic potential at zero temperature from that at finite temperature}

In this appendix we derive the thermodynamic potential at zero temperature from that at finite temperature.
The thermodynamic potential at finite temperature is as follows:
\begin{align}
\varOmega(M,F,\mu,T)
&=-\sum_{\eta,\tau,\alpha}\int_{\vec{p}^2\le \varLambda^2} \frac{d^3p}{(2\pi)^3}
\Biggl\{ E^{(\eta)}_{\vec{p}}+T \log \left( 1+\exp \left( -\frac{E^{(\eta)}_{\vec{p}}-\mu}{T} \right) \right)\nonumber\\
&\qquad\qquad\qquad\qquad\qquad\qquad\ \ 
+T \log \left( 1+\exp \left( -\frac{E^{(\eta)}_{\vec{p}}+\mu}{T} \right) \right) \Biggl\} \nonumber\\
&+\frac{F^2}{2G_{\text{T}}}+\frac{M^2}{4G_{\text{S}}}.
\end{align}
If we assume $T \ll 1$, we can carry out the Taylor expansion for the logarithmic function in the following way:
\begin{equation*}
T \log \left( 1+\exp \left( -\frac{E^{(\eta)}_{\vec{p}}-\mu}{T} \right) \right) \rightarrow
\begin{cases}
-(E^{(\eta)}_{\vec{p}}-\mu)+T \exp \left( \frac{E^{(\eta)}_{\vec{p}}-\mu}{T} \right) & \text{for}\  E^{(\eta)}_{\vec{p}} \le \mu \\
T \exp \left( -\frac{E^{(\eta)}_{\vec{p}}-\mu}{T} \right) & \text{for}\  E^{(\eta)}_{\vec{p}}>\mu,
\end{cases}
\end{equation*}
\begin{equation}
T \log \left( 1+\exp \left( -\frac{E^{(\eta)}_{\vec{p}}+\mu}{T} \right) \right) \rightarrow T \exp \left( -\frac{E^{(\eta)}_{\vec{p}}+\mu}{T} \right).
\qquad\qquad\qquad\qquad\qquad\ \ 
\end{equation}
Furthermore, in the region where $T \rightarrow 0$ we can reduce the above expressions into
\begin{equation*}
T \log \left( 1+\exp \left( -\frac{E^{(\eta)}_{\vec{p}}-\mu}{T} \right) \right) \rightarrow
-(E^{(\eta)}_{\vec{p}}-\mu) \, \theta(\mu-E^{(\eta)}_{\vec{p}}),
\end{equation*}
\begin{equation}
T \log \left( 1+\exp \left( -\frac{E^{(\eta)}_{\vec{p}}+\mu}{T} \right) \right) \rightarrow 0, 
\qquad\qquad\qquad\qquad\qquad
\end{equation}
where $\theta(x)$ is the step function.
Using these results, we could rewrite $\varOmega(M,F,\mu,T)$ into
\begin{align}
\varOmega(M,F,\mu,T) \rightarrow &-\sum_{\eta,\tau,\alpha} \int_{\vec{p}^2 \le \varLambda^2} \frac{d^3p}{(2\pi)^3}
\left\{ E^{(\eta)}_{\vec{p}}-(E^{(\eta)}_{\vec{p}}-\mu) \, \theta(\mu-E^{(\eta)}_{\vec{p}}) \right\} \nonumber\\
&\ \ +\frac{F^2}{2G_{\text{T}}}+\frac{M^2}{4G_{\text{S}}}.
\end{align}
This expression is just one for the thermodynamic potential at zero temperature.

\setcounter{equation}{0}

\section{Derivation for the effective potential with functional method}

Let us start with the following Lagrangian density in order to derive the effective potential by using the functional method:
\begin{align}
\mathcal{L}&=\bar{\psi}(i \gamma^\mu \partial_\mu - m_0)\psi + G_S (\bar{\psi}\psi)^2 - 
\frac{G_T}{2} (\bar{\psi}\gamma^1 \gamma^2 \tau_3 \psi)(\bar{\psi}\gamma_1 \gamma_2 \tau_3 \psi) \notag \\
&=\bar{\psi}(i \gamma^\mu \partial_\mu - m_0)\psi + G_S (\bar{\psi}\psi)^2 + \frac{G_T}{2} (\bar{\psi}\varSigma_3 \psi)^2,
\end{align}
where we define $\varSigma_3 \equiv -i \gamma^1 \gamma^2$. 
In order to perform functional integral, let us introduce two auxiliary fields, $M'$ and $F$, and use a relation of unit:
\begin{align}
1=\int \mathcal{D}M' \mathcal{D}F &\exp \left[ -i \int d^4x \left\{ M'+G_S (\bar{\psi}\psi) \right\} G^{-1}_{S} \left\{ M'+G_S (\bar{\psi}\psi) \right\}\right] 
\notag \\
&\times \exp \left[ -\frac{i}{2} \int d^4x \left\{ F+G_T (\bar{\psi}\varSigma_3 \psi) \right\} G^{-1}_{T} \left\{ F+G_T (\bar{\psi}\varSigma_3 \psi) 
\right\}\right].
\end{align}
The generating functional $Z$ for the Lagrangian density (B1) is given as follows:
\begin{equation}
Z \propto \int \mathcal{D}\bar{\psi} \mathcal{D}\psi \exp \left[ i\int d^4x \left\{ \bar{\psi}(i \gamma^\mu \partial_\mu - m_0)\psi
+ G_S (\bar{\psi}\psi)^2 + \frac{G_T}{2}(\bar{\psi}\varSigma_3 \psi)^2 \right\} \right].
\end{equation}
Inserting the relation of unit, (B2), into $Z$ and setting $M' =M/2$, we obtain
\begin{equation}
Z \propto \int \mathcal{D}\bar{\psi} \mathcal{D}\psi \mathcal{D}M \mathcal{D}F
\exp \left[ i\int d^4x \left\{ \bar{\psi}(i \gamma^\mu \partial_\mu-M_q-F \varSigma_3)\psi-\frac{M^2}{4G_S}-\frac{F^2}{2G_T} \right\} \right],
\end{equation}
where we define $M_q \equiv m_0+M$. 
Thus, we can integrate out with respect to $\psi$ and $\bar{\psi}$ easily.
After some calculations, we get
\begin{align}
Z &\propto \int \mathcal{D}M \mathcal{D}F \; \text{Det}(i\gamma^\mu \partial_\mu-M_q-F \varSigma_3) 
\exp \left[ -i\int d^4x \left( \frac{M^2}{4G_S}+\frac{F^2}{2G_T} \right) \right] \notag \\
&=\int \mathcal{D}M \mathcal{D}F \exp \left[ \text{Tr} \log \det (i\gamma^\mu \partial_\mu-M_q-F \varSigma_3)
-i\int d^4x \left( \frac{M^2}{4G_S}+\frac{F^2}{2G_T} \right) \right],
\end{align}
where in the second line the determinant, $\det$, operates on gamma matrices. 
In order to compute the trace, Tr, we change to momentum space.
\begin{align}
Z \propto \int \mathcal{D}M \mathcal{D}F &\exp \bigg[ iN_C N_F \int d^4x \frac{d^4p}{i(2\pi)^4} \log \det (\not p -M_q-F\varSigma_3) \bigg] \notag \\
&\times \exp \bigg[ -i\int d^4x \left( \frac{M^2}{4G_S}+\frac{F^2}{2G_T} \right) \bigg],
\end{align}
where $N_C$ and $N_F$ mean the number of color and flavor, respectively.
Our next step is to calculate the determinant.
We can do it as follows:
\begin{align}
&\det(\not p -M_q-F\varSigma_3) \notag \\
&=\det \gamma^0 (\not p -M_q-F\varSigma_3) \notag \\
&=\det \left[ p^0-\gamma^0 (\vec{\gamma} \cdot \vec{p}+M_q+F \varSigma_3) \right] \notag \\
&=\det \left[ p^0 -\begin{pmatrix}
E^{(+)}_{\vec{p}} & & & \\
&E^{(-)}_{\vec{p}} & & \\
& & -E^{(+)}_{\vec{p}} & \\
& & & -E^{(-)}_{\vec{p}}
\end{pmatrix} \right] \notag \\
&=(p^0-E^{(+)}_{\vec{p}})(p^0-E^{(-)}_{\vec{p}})(p^0+E^{(+)}_{\vec{p}})(p^0+E^{(-)}_{\vec{p}})\ .
\end{align}
Substituting the above result into $Z$ in (B6), we obtain 
\begin{align}
Z \propto \int &\mathcal{D}M \mathcal{D}F \notag \\
&\times \exp \left[ iN_C N_F \int d^4x \frac{d^4 p}{i(2\pi)^4} 
\log (p^0-E^{(+)}_{\vec{p}})(p^0-E^{(-)}_{\vec{p}})(p^0+E^{(+)}_{\vec{p}})(p^0+E^{(-)}_{\vec{p}}) \right] \notag \\
&\times \exp \left[ -i \int d^4 x \left( \frac{M^2}{4G_S}+\frac{F^2}{2G_T} \right) \right].
\end{align}
For a little while, we consider the only contents of exponential in the second line in (B8):
\begin{equation}
\int \frac{d^4 p}{i(2\pi)^4} \log (p^0-E^{(+)}_{\vec{p}})(p^0-E^{(-)}_{\vec{p}})(p^0+E^{(+)}_{\vec{p}})(p^0+E^{(-)}_{\vec{p}}).
\end{equation}
Let us differentiate and integrate the above expression with respect to $E^{(+)}_{\vec{p}}$ and $E^{(-)}_{\vec{p}}$.
As a result, (B9) can be recast into 
\begin{align}
\int \frac{d^4 p}{i(2\pi)^4} &\int d E^{(+)}_{\vec{p}} \left( \frac{1}{p^0+E^{(+)}_{\vec{p}}}-\frac{1}{p^0-E^{(+)}_{\vec{p}}} \right) \notag \\
&+\int \frac{d^4 p}{i(2\pi)^4} \int d E^{(-)}_{\vec{p}} \left( \frac{1}{p^0+E^{(-)}_{\vec{p}}}-\frac{1}{p^0-E^{(-)}_{\vec{p}}} \right).
\end{align}
We would like to discuss a system at finite temperature and density.
So let us change the integration to the summation by using the Matsubara method as follows:
\begin{equation}
\int \frac{d^4 p}{i(2\pi)^4} f(p^0,\vec{p}) \rightarrow T \sum_{n=-\infty}^{\infty} \int \frac{d^3 p}{(2\pi)^3} f(i\omega_n +\mu,\vec{p}),
\end{equation}
where $\omega_n$ is the Matsubara frequency and $\mu$ is chemical potential.
Using a formula 
\begin{equation}
\lim_{\epsilon \rightarrow +0} T\sum_n \frac{e^{i \omega_n \epsilon}}{i \omega_n -x}
=\lim_{\epsilon \rightarrow +0} \frac{e^{i \omega_n \epsilon}}{e^{x/T}+1}
=\frac{1}{e^{x/T}+1},
\end{equation}
we can calculate the summation following the standard technique.
As a result,  we obtain
\begin{eqnarray}
& &\sum_{\eta=\pm} \int \frac{d^3 p}{(2\pi)^3} \left[ E^{(\eta)}_{\vec{p}} +\mu+T \log \left\{ 1+\exp \left( -\frac{E^{(\eta)}_{\vec{p}} +\mu}{T} \right) \right\}
\right.
\nonumber\\
& &\qquad\qquad\qquad\qquad\qquad\left.
+T \log \left\{ 1+\exp \left( -\frac{E^{(\eta)}_{\vec{p}}-\mu}{T} \right) \right\} \right].
\end{eqnarray}
Substituting the above result into $Z$, $Z$ can be expressed as
\begin{align}
Z \propto \int \mathcal{D}M \mathcal{D}F \exp i&\bigg[ N_C N_F \int d^4 x \frac{d^3 p}{(2\pi)^3} \sum_{\eta} 
\bigg( E^{(\eta)}_{\vec{p}}+T \log \bigg\{ 1+\exp \bigg( -\frac{E^{(\eta)}_{\vec{p}}+\mu}{T} \bigg) \bigg\} \notag \\
&+T \log \bigg\{ 1+\exp \bigg( -\frac{E^{(\eta)}_{\vec{p}}-\mu}{T} \bigg) \bigg\}\bigg)-\int d^4x \bigg( \frac{M^2}{4G_S}+\frac{F^2}{2G_T} \bigg) \bigg],
\end{align}
where we neglect a constant term.
In general, the effective action $\varGamma$ and effective potential $V$ are defined as follows:
\begin{equation}
Z=\exp \left( i \varGamma[M,F,T,\mu] \right),\ \qquad
V[M,F,T,\mu]=-\frac{\varGamma[M,F,T,\mu]}{\int d^4 x}.
\end{equation}
Thus, finally, we obtain the effective potential $V$ as 
\begin{align}
V[M,F,T,\mu]= &-N_C N_F \int \frac{d^3 p}{(2\pi)^3} \sum_{\eta} 
\bigg[ E^{(\eta)}_{\vec{p}}+T \log \bigg\{ 1+\exp \bigg( -\frac{E^{(\eta)}_{\vec{p}}+\mu}{T} \bigg) \bigg\} \notag \\
&+T \log \bigg\{ 1+\exp \bigg( -\frac{E^{(\eta)}_{\vec{p}}-\mu}{T} \bigg) \bigg\}\bigg]+\frac{M^2}{4G_S}+\frac{F^2}{2G_T}.
\end{align}
This is identical with the thermodynamic potential (\ref{5-2}).

\setcounter{equation}{0}

\section{The domain of integral with respect to the three-momentum in the thermodynamic potential at zero temperature}

In {\bf 3}, we gave the expression of the thermodynamic potential. 
In this appendix, we give a domain of integral with respect to three-momentum carefully.  
We assume $M \ge 0$, $F \ge 0$ and $\varLambda > \mu$ without loss of generality and introduce polar coordinates ($p_T,\theta$):
\[
p_1=p_T \cos \theta,\, \qquad
p_2=p_T \sin \theta.
\]
Moreover, we define $q \equiv \sqrt{p_T^2+M_{\text{q}}^2}$ in order to integrate over momenta.
After integrating over $\theta$ in Eqs.(3.4) - (3.6), we obtain the thermodynamic potential $\varPhi=\sum_{i=1}^{4}\varPhi_i$ as follows:
\begin{align*}
\varPhi_1(M,F,\mu)=&\frac{3}{2\pi^2} \int_{\varGamma_1} dq dp_3 \, q\left( \sqrt{p_3^2+(q+F)^2}-\mu \right), \notag \\
&\varGamma_1=\left\{ p_3^2+(q+F)^2 \le \mu^2,p_3^2+q^2 \le \varLambda^2+M^2_{\text{q}},q \ge M_{\text{q}}\right\}
\end{align*}
\begin{align*}
\varPhi_2(M,F,\mu)=&\frac{3}{2\pi^2} \int_{\varGamma_2} dq dp_3 \, q\left( \sqrt{p_3^2+(q-F)^2}-\mu \right), \notag \\
&\varGamma_2=\left\{ p_3^2+(q-F)^2 \le \mu^2,p_3^2+q^2 \le \varLambda^2+M^2_{\text{q}},q \ge M_{\text{q}}\right\}
\end{align*}
\begin{align*}
\varPhi_3(M,F,\mu)=&-\frac{3}{2\pi^2} \sum_{\eta} \int_{\varGamma_3} dq dp_3 \, q\sqrt{p_3^2+(q+\eta F)^2}, \qquad\qquad\qquad \notag \\
&\varGamma_3=\left\{ p_3^2+q^2 \le \varLambda^2+M^2_{\text{q}},q \ge M_{\text{q}} \right\}
\end{align*}
\begin{equation*}
\varPhi_4(M,F,\mu)=\frac{M^2}{4G_{\text{S}}}+\frac{F^2}{2G_{\text{T}}}.\qquad\qquad\qquad\qquad\qquad\qquad\qquad\qquad\quad
\end{equation*}
Further, let us integrate the thermodynamic potential over $p_3$ analytically.
To do this we must discuss the domain of the integral carefully.
First we consider $\varPhi_1$ and $\varGamma_1$.
If the first condition in $\varGamma_1$ is satisfied, the second condition in it will be satisfied automatically.
So we could reduce $\varGamma_1$ to
\[
\varGamma_1=\left\{ p_3^2+(q+F)^2 \le \mu^2,q \ge M_{\text{q}} \right\}.
\]
Furthermore, we could change the above condition to as follows:
\[
\varGamma_1=\left\{ -\sqrt{\mu^2-(q+F)^2} \le p_3 \le \sqrt{\mu^2-(q+F)^2},\; q \ge M_{\text{q}} \right\}.
\]
Since the contents in a square root must be positive, the final form is
\begin{equation*}
\varGamma_1=\left\{ -\sqrt{\mu^2-(q+F)^2} \le p_3 \le \sqrt{\mu^2-(q+F)^2},\; M_{\text{q}} \le q \le \mu-F \right\}.
\end{equation*}
However, we need the condition: $M_{\text{q}} \le \mu-F$ in order to integrate over $q$.
If $M_{\text{q}} > \mu-F$, we can not perform integral.
Using an integration formula:
\[
\int dx \, \sqrt{x^2+a^2} = \frac{1}{2} \left\{ x\sqrt{x^2+a^2}+a^2\log \left( x+\sqrt{x^2+a^2} \right) \right\},
\]
we were able to perform integral over $p_3$ easily.
The final results are the following:
\bsub\label{c1}
\beq
& &{\rm If}\  M_{\text{q}} > \mu-F, \nonumber\\
& &\quad
\varPhi_1(M,F,\mu)=0.
\label{c1a}\\
& &
{\rm If}\ M_{\text{q}} \le \mu-F, \nonumber\\
& &\quad
\varPhi_1(M,F,\mu)=\frac{3}{2\pi^2}\int_{M_{\text{q}}}^{\mu-F} dq\, q\biggl\{ -\mu\sqrt{\mu^2-(q+F)^2}
\nonumber\\
& &\qquad\qquad\qquad\qquad\qquad\qquad\quad
+(q+F)^2
\log \left( \frac{\sqrt{\mu^2-(q+F)^2}+\mu}{q+F} \right) \biggl\}.\ \ 
\label{c1b}
\eeq
\esub

Next, we consider $\varPhi_2$ and $\varGamma_2$.
This case is more complicated than the previous case.
There are five cases for conditions to perform integral as follows:
\begin{flushleft}
If $F-\mu \le M_{\text{q}} \le F+\mu \le \sqrt{\varLambda^2+M^2_{\text{q}}}$,
\end{flushleft}
\begin{equation*}
\varPhi_2(M,F,\mu)=\frac{3}{2\pi^2}\int_{M_{\text{q}}}^{F+\mu}dq \int_{-\sqrt{\mu^2-(q-F)^2}}^{\sqrt{\mu^2-(q-F)^2}}dp_3\, 
q\left( \sqrt{p_3^2+(q-F)^2}-\mu \right).
\end{equation*}
\begin{flushleft}
If $M_{\text{q}} \le F-\mu \le F+\mu \le \sqrt{\varLambda^2+M^2_{\text{q}}}$,
\end{flushleft}
\begin{equation*}
\varPhi_2(M,F,\mu)=\frac{3}{2\pi^2}\int_{F-\mu}^{F+\mu}dq \int_{-\sqrt{\mu^2-(q-F)^2}}^{\sqrt{\mu^2-(q-F)^2}}dp_3\, 
q\left( \sqrt{p_3^2+(q-F)^2}-\mu \right).
\end{equation*}
\begin{flushleft}
If $M_{\text{q}} \le F-\mu \le b \le \sqrt{\varLambda^2+M^2_{\text{q}}} \le F+\mu$,
\end{flushleft}
\begin{align*}
\varPhi_2(M,F,\mu)&=\frac{3}{2\pi^2}\left( \int^{b}_{F-\mu} \int_{-\sqrt{\mu^2-(q-F)^2}}^{\sqrt{\mu^2-(q-F)^2}}+
\int_b^{\sqrt{\varLambda^2+M^2_{\text{q}}}} \int_{-\sqrt{\varLambda^2+M^2_{\text{q}}-q^2}}^{\sqrt{\varLambda^2+M^2_{\text{q}}-q^2}} \right) \notag \\
&\times q\left( \sqrt{p_3^2+(q-F)^2}-\mu \right) dp_3\, dq.
\end{align*}
\begin{flushleft}
If $F-\mu \le M_{\text{q}} \le b \le \sqrt{\varLambda^2+M^2_{\text{q}}} \le F+\mu$,
\end{flushleft}
\begin{align*}
\varPhi_2(M,F,\mu)&=\frac{3}{2\pi^2}\left( \int^{b}_{M_{\text{q}}} \int_{-\sqrt{\mu^2-(q-F)^2}}^{\sqrt{\mu^2-(q-F)^2}}+
\int_b^{\sqrt{\varLambda^2+M^2_{\text{q}}}} \int_{-\sqrt{\varLambda^2+M^2_{\text{q}}-q^2}}^{\sqrt{\varLambda^2+M^2_{\text{q}}-q^2}} \right) \notag \\
&\times q\left( \sqrt{p_3^2+(q-F)^2}-\mu \right) dp_3\, dq.
\end{align*}
\begin{flushleft}
If $F-\mu \le b \le M_{\text{q}} \le \sqrt{\varLambda^2+M_{\text{q}}} \le F+\mu$,
\end{flushleft}
\begin{equation*}
\varPhi_2(M,F,\mu)=\frac{3}{2\pi^2} \int_{M_{\text{q}}}^{\sqrt{\varLambda^2+M^2_{\text{q}}}} dq 
\int_{-\sqrt{\varLambda^2+M^2_{\text{q}}-q^2}}^{\sqrt{\varLambda^2+M^2_{\text{q}}-q^2}}dp_3\,q\left( \sqrt{p_3^2+(q-F)^2}-\mu \right).
\end{equation*}
Here, $b$ is the solution for $q$ of the simultaneous equation:
\begin{equation*}
p_3^2+q^2=\varLambda^2+M^2_{\text{q}} \ , \qquad  
p_3^2+(q-F)^2=\mu^2
\end{equation*}
%The solution is
%\begin{equation*}
%q=\frac{\varLambda^2+M^2_{\text{q}}+F^2-\mu^2}{2F}\equiv b.
%\end{equation*}
We were able to perform integral over $p_3$, then define two functions for simplicity as follows:
\begin{equation*}
\phi_1(q) \equiv \frac{3}{2\pi^2}q \left\{ -\mu \sqrt{\mu^2-(F-q)^2}+(F-q)^2
\log \left( \frac{\mu+\sqrt{\mu^2-(F-q)^2}}{|F-q|} \right) \right\}, \ \ 
\end{equation*}
\begin{align*}
\phi_2(q) &\equiv \frac{3}{2\pi^2}q \Biggl\{ \sqrt{\varLambda^2+M^2_{\text{q}}-q^2} 
\left( -2\mu+\sqrt{F^2+\varLambda^2+M^2_{\text{q}}-2Fq} \right) \\
&\qquad \qquad 
+(F-q)^2 \log \left( \frac{\sqrt{F^2+\varLambda^2+M^2_{\text{q}}-2Fq}+\sqrt{\varLambda^2+M^2_{\text{q}}-q^2}}{|F-q|} \right) \Biggr\}.
\end{align*}
Using the above expressions, the final results are summarized as follows:
\bsub\label{c2}
\beq
& &
{\rm If}\ F-\mu \le M_{\text{q}} \le F+\mu \le \sqrt{\varLambda^2+M^2_{\text{q}}}, \nonumber\\
& &\qquad
\varPhi_2(M,F,\mu)=\int_{M_{\text{q}}}^{F+\mu}dq\, \phi_1(q).
\label{c2a}\\
& &{\rm If}\ M_{\text{q}} \le F-\mu \le F+\mu \le \sqrt{\varLambda^2+M^2_{\text{q}}},\nonumber\\
& &\qquad
\varPhi_2(M,F,\mu)=\int_{F-\mu}^{F+\mu}dq\, \phi_1(q).
\label{c2b}\\
& &{\rm If}\ M_{\text{q}} \le F-\mu \le b \le \sqrt{\varLambda^2+M^2_{\text{q}}} \le F+\mu,\nonumber\\
& &\qquad
\varPhi_2(M,F,\mu)=\int^{b}_{F-\mu} dq\, \phi_1(q)+\int_{b}^{\sqrt{\varLambda^2+M^2_{\text{q}}}} dq\, \phi_2(q).
\label{c2c}\\
& &{\rm If}\ F-\mu \le M_{\text{q}} \le b \le \sqrt{\varLambda^2+M^2_{\text{q}}} \le F+\mu,\nonumber\\
& &\qquad
\varPhi_2(M,F,\mu)=\int^{b}_{M_{\text{q}}}dq\,\phi_1(q)+\int_b^{\sqrt{\varLambda^2+M^2_{\text{q}}}}dq\,\phi_2(q).
\label{c2d}\\
& &{\rm If}\ F-\mu \le b \le M_{\text{q}} \le \sqrt{\varLambda^2+M_{\text{q}}} \le F+\mu,\nonumber\\
& &\qquad
\varPhi_2(M,F,\mu)=\int_{M_{\text{q}}}^{\sqrt{\varLambda^2+M^2_{\text{q}}}}d\,q \phi_2(q).
\label{c2e}
\eeq
\esub

Finally, we discuss $\varPhi_3$ and $\varGamma_3$.
We could derive the domain of integral easily in this case.
The domain is re-expressed as
\begin{equation*}
\varGamma_3=\left\{ -\sqrt{\varLambda^2+M^2_{\text{q}}-q^2} \le p_3 \le \sqrt{\varLambda^2+M^2_{\text{q}}-q^2},\  
M_{\text{q}} \le q \le \sqrt{\varLambda^2+M^2_{\text{q}}} \right\}.
\end{equation*}
After integrating over $p_3$, we obtain the final result:
\begin{align}
\varPhi_3(M,F,\mu)&=-\frac{3}{2\pi^2} \sum_{\eta} \int_{M_{\text{q}}}^{\sqrt{\varLambda^2+M^2_{\text{q}}}} dq\,
q\Biggl\{ \sqrt{\varLambda^2+M^2_{\text{q}}-q^2}\sqrt{\varLambda^2+M^2_{\text{q}}+2\eta qF+F^2} \notag \\
&+(q+\eta F)^2\log \left( \frac{\sqrt{\varLambda^2+M^2_{\text{q}}-q^2}+\sqrt{\varLambda^2+M^2_{\text{q}}+2\eta qF+F^2}}{|q+\eta F|} \right) \Biggr\}.
\end{align}

\setcounter{equation}{0}

\section{A possibility of origin of tensor-type interaction in NJL model}

In the QCD Lagrangian, the interaction part is written as 
\beq\label{d1}
{\cal L}_{\rm int}=g{\bar \psi}(x)\gamma^\mu \psi(x)A_\mu(x)\ ,
\eeq
where $\psi(x)$ and $A_\mu(x)\ (=A_\mu^a(x) T^a)$ are quark and gluon fields, respectively, and 
$T^a$ represents the color $su(3)$-generators. 
%In this appendix, $T^a$ has no essential role for the purpose to derive the four-point interaction.  
Here, two-gluon exchange diagram is properly depicted in Fig.{\ref{fig:a1}}. 
These diagrams are obtained from fourth-order perturbation of ${\cal L}_{\rm int}$:
\beq\label{d2}
& &\int d^4x{\cal L}_{\rm int}(x)\cdot\int d^4y{\cal L}_{\rm int}(y)\cdot
\int d^4x{\cal L}_{\rm int}(x')\cdot\int d^4x{\cal L}_{\rm int}(y')
\nonumber\\
&=& 
g^4\int d^4x d^4y d^4x' d^4y'
{\bar \psi}(x)\gamma^\mu \psi(x)A_\mu(x)
{\bar \psi}(y)\gamma^\nu \psi(y)A_\nu(y)
\nonumber\\
& &\qquad\qquad\qquad\qquad
\times
{\bar \psi}(x')\gamma^\rho \psi(x')A_\rho(x')
{\bar \psi}(y')\gamma^\sigma \psi(y')A_\sigma(y')\ .\qquad
\eeq
We here intend to describe the above expression as 
\beq\label{d3}
\int d^4x {\cal L}_{\rm eff}\ , 
\eeq
which should be expressed as the four-point interaction between quarks.

%
%%%%%%%%%%%%%%%%%%%%%%%%%%%%%%%%%%%%%%%%%%%%%%%%%%%%%%%%%%%%%%%%%%%%%%
\begin{figure}[b]
\begin{center}
\includegraphics[height=4.0cm]{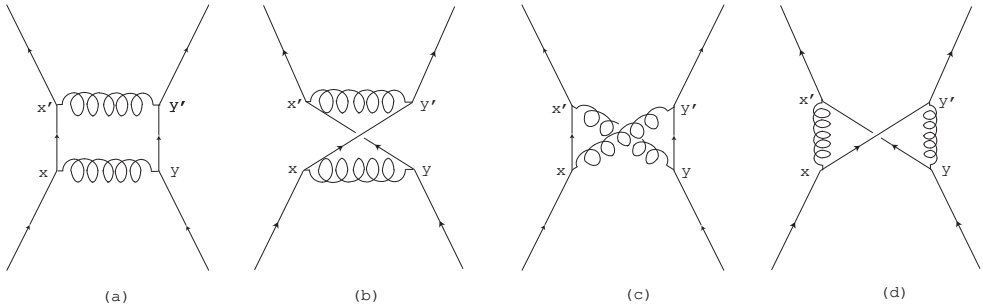}
\caption{
The Feynman diagrams of proper two-gluon exchange process. 
}
\label{fig:a1}
\end{center}
\end{figure}
%%%%%%%%%%%%%%%%%%%%%%%%%%%%%%%%%%%%%%%%%%%%%%%%%%%%%%%%%%%%%%%%%%%%%%%%
%

As for the process in Fig.{\ref{fig:a1}} (a) and (b), 
these process may mainly be 
regarded as the repeated processes of one-gluon exchange process.
Thus, we omit these process in proper contribution of two-gluon exchange process.
Therefore, let us first consider the diagram in Fig.{\ref{fig:a1}} (c). 
Writing the spinor indices, $i, j, \cdots$, explicitly, we contract bilinear field making the 
Feynman propagator:
\beq\label{d4}
& &
g^4\int d^4x d^4y d^4x' d^4y'
{\bar \psi}_i(x)\gamma^\mu_{ij }{\psi}_j(x)A_\mu(x){\bar \psi}_k(y)\gamma^\nu_{kl} \psi_l(y)A_\nu(y)
\nonumber\\
& &\qquad\qquad\qquad\qquad
\times
{{\bar \psi}}_m(x') \gamma^\rho_{mn} \psi_n(x')A_\rho(x')
{\bar \psi}_p(y')\gamma^\sigma_{pq} \psi_q(y')A_\sigma(y')
\nonumber\\
& &\longrightarrow 
g^4\int d^4x d^4y d^4x' d^4y' {\cal L}_{(c)}\ , \nonumber\\
& &{\cal L}_{(c)}
=-{\bar \psi}_i(x)\gamma^{\mu}_{ij}{\bar \psi}_k(y)\gamma^{\nu}_{kl}\gamma^{\rho}_{mn}
\psi_n(x')\gamma^{\sigma}_{pq}\psi_q(y')
\nonumber\\
& &\qquad\qquad\qquad\qquad
\times
\langle \psi_j(x){\bar \psi}_m(x')\rangle
\langle A_\mu(x)A_{\sigma}(y')\rangle
\langle \psi_l(y){\bar \psi}_p(y')\rangle
\langle A_\nu(y)A_{\rho}(x')\rangle .
\eeq
Here, it should be noted that the property of Grassmann number for fermion field 
is used. Thus, a minus sign appears. 
Here, $\langle \psi_i(x){\bar \psi}_j(y)\rangle$ and 
$\langle A_\mu(x)A_{\nu}(y)\rangle\ (=T^a T^b \langle A_\mu^a(x)A_{\nu}^b(y)\rangle)$ represent the Feynman propagator for 
quark and gluon fields, respectively: 
\beq\label{d5}
& &\langle \psi_i(x){\bar \psi}_j(y)\rangle
=\int \frac{d^4 p}{i(2\pi)^4}\frac{\gamma^\mu p_\mu +M_q}{M_q^2-p^2-i\epsilon}e^{-ip(x-y)}\ , 
\nonumber\\
& &
\langle A_\mu^a(x)A_{\nu}^b(y)\rangle
=\delta^{ab}\int \frac{d^4 p}{i(2\pi)^4}
\frac{1}{p^2+i\epsilon}
\left[g_{\mu\nu}-(1-\alpha)\frac{p_\mu p_\nu}{p^2}\right]e^{-ip(x-y)}\ , 
\eeq
where $a$ and $b$ are color indices, $M_q$ is the quark mass and $\alpha$ is a gauge parameter.
Of course, the NJL model Lagrangian cannot be derived from QCD. 
Therefore, we have to give up the exact calculation. 
Thus, we assume the form of propagators so as to reproduce the four-point 
contact interaction between quarks.
As for the quark propagator, the quark mass in the propagator is set to very large 
value or infinity artificially:
\beq\label{quark}
& &\langle \psi_i(x){\bar \psi}_j(y)\rangle
\sim
\int \frac{d^4 p}{i(2\pi)^4}\frac{1_{ij}}{M_q}e^{-ip(x-y)}=\frac{1}{iM_q}\delta_{ij}\delta^4(x-y)\ .
\eeq
As for the gluon propagator, artificially ``gluon mass" $M_g$ is introduced and 
is taken as a very large value or  infinity.
\beq\label{gluon}
\langle A_\mu^a(x)A_{\nu}^b(y)\rangle
&=&\delta^{ab}\int \frac{d^4 p}{i(2\pi)^4}
\frac{1}{p^2+i\epsilon}
\left[g_{\mu\nu}-(1-\alpha)\frac{p_\mu p_\nu}{p^2}\right]e^{-ip(x-y)}
\nonumber\\
&\rightarrow&
\delta^{ab}\int \frac{d^4 p}{i(2\pi)^4}
\frac{1}{p^2-M_g^2+i\epsilon}
\left[g_{\mu\nu}-(1-\alpha)\frac{p_\mu p_\nu}{p^2}\right]e^{-ip(x-y)}
\nonumber\\
&\sim&
\delta^{ab}\int \frac{d^4 p}{i(2\pi)^4}
\frac{g_{\mu\nu}}{-M_g^2}e^{-ip(x-y)}
=-\frac{1}{iM_g^2}\delta^{ab}g_{\mu\nu}\delta^4(x-y) . 
\eeq
Hereafter, we denote $M_q^2M_g^4\equiv M_{\rm eff}^6$, in which $M_{\rm eff}$ has mass-dimension.
Inserting the above ``approximate" propagators into Eq.(\ref{d4}), then, 
Eq.(\ref{d4}) is rewritten as 
\beq\label{d8}
& &{\cal L}_{(c)}
=\frac{C_2^2}{M_{\rm eff}^6}
{\bar \psi}_i(x)\gamma^{\mu}_{ij}\gamma^{\rho}_{jn}\psi_n(x)
{\bar \psi}_k(x)\gamma_{\rho, kl}\gamma_{\mu,lq}\psi_q(x) \nonumber\\
& &\qquad\qquad\qquad\qquad\qquad
\times
\delta^4(x-x')\delta^4(x'-y)\delta^4(x-y')\delta^4(y'-y) ,
\eeq
where $C_2=\sum_a T^a T^a$. 
Here, we use again the property of Grassmann number. 
Thus, we obtain 
\beq\label{d9}
g^4\int d^4x d^4y d^4x' d^4 y' {\cal L}_{(c)}
&=&\frac{g^4\delta^4(0)}{M_{\rm eff}^6}\cdot C_2^2
\int d^4 x\ {\bar \psi}(x)\gamma^\mu \gamma^{\rho}\psi(x)\cdot {\bar \psi}(x)
\gamma_{\rho}\gamma_{\mu}\psi(x)\nonumber\\
&=&-g_T\int d^4 x {\bar \psi}(x)\gamma^\mu \gamma^{\nu}\psi(x)\cdot {\bar \psi}(x)
\gamma_{\mu}\gamma_{\nu}\psi(x)
\nonumber\\
& &
+8g_T\int d^4 x{\bar \psi}(x)\psi(x)\cdot {\bar \psi}(x)\psi(x)\ , 
\eeq
where we define 
$g_T=g^4\delta^4(0)C_2^2/M_{\rm eff}^6\ (>0)$ and use $\gamma^\mu\gamma^\nu+\gamma^\nu\gamma^\mu=2g^{\mu\nu}$ 
and $\gamma^\mu \gamma_\mu=4$. 
Here, $\delta^4(0)=\int d^4k/(2\pi)^4\cdot \left. e^{ik(x-y)}\right|_{x=y}$ which is regarded as a very large value or infinity 
in order to $g_T$ has a finite value. 
Then, $g_T$ has a dimension of  (mass)$^{-2}$.

Next, let us consider Fig.{\ref{fig:a1} (d).
As is similar to the case Fig.{\ref{fig:a1} (c), we obtain 
\beq\label{d10}
& &
g^4\int d^4x d^4y d^4x' d^4y'
{\bar \psi}_i(x)\gamma^\mu_{ij} {\psi}_j(x) A_\mu(x)
{\bar \psi}_k(y)\gamma^\nu_{kl} \psi_l(y)A_\nu(y)
\nonumber\\
& &\qquad\qquad\qquad\qquad
\times
{{\bar \psi}}_m(x') \gamma^\rho_{mn} \psi_n(x')A_\rho(x')
{\bar \psi}_p(y')\gamma^\sigma_{pq} \psi_q(y')A_\sigma(y')
\nonumber\\
& &\longrightarrow
g^4\int d^4x d^4y d^4x' d^4y' {\cal L}_{(d) \ , }\nonumber\\
& &{\cal L}_{(d)}
={\bar \psi}_i(x)\gamma^{\mu}_{ij}{\bar \psi}_k(y)\gamma^{\nu}_{kl}\gamma^{\rho}_{mn}
\psi_n(x')\gamma^{\sigma}_{pq}\psi_q(y')
\nonumber\\
& &\qquad\qquad\qquad\qquad
\times
\langle \psi_l(y){\bar \psi}_m(x')\rangle
\langle A_\mu(x)A_{\rho}(x')\rangle
\langle \psi_j(x){\bar \psi}_p(y')\rangle
\langle A_\nu(y)A_{\sigma}(y')\rangle . \quad
\eeq
In order to obtain the four-point contact interaction for NJL type, 
we ``approximate" the propagators in (\ref{quark}) and (\ref{gluon}). 
Then,
\beq\label{d11}
& &{\cal L}_{(d)}
=\frac{C_2^2}{M_{\rm eff}^6}
{\bar \psi}_i(x)\gamma^{\mu}_{ij}\gamma^{\sigma}_{jq}\psi_q(x)
{\bar \psi}_k(x)\gamma_{\sigma, kl}\gamma_{\mu,ln}\psi_n(x)
\nonumber\\
& &\qquad\qquad\qquad\qquad
\times
\delta^4(x-x')\delta^4(x'-y)\delta^4(x-y')\delta^4(y'-y)
\nonumber\\
& &
\eeq
is obtained. 
Therefore, 
\beq\label{d12}
g^4\int d^4x d^4y d^4x' d^4 y' {\cal L}_{(d)}
&=&\frac{g^4\delta^4(0)}{M_{\rm eff}^6}\cdot C_2^2
\int d^4 x\ {\bar \psi}(x)\gamma^\mu \gamma^{\nu}\psi(x)\cdot {\bar \psi}(x)
\gamma_{\nu}\gamma_{\mu}\psi(x)\nonumber\\
&=&-g_T\int d^4 x {\bar \psi}(x)\gamma^\mu \gamma^{\nu}\psi(x)\cdot {\bar \psi}(x)
\gamma_{\mu}\gamma_{\nu}\psi(x)
\nonumber\\
& &
+8g_T\int d^4 x{\bar \psi}(x)\psi(x)\cdot {\bar \psi}(x)\psi(x)\ , 
\eeq

%
%%%%%%%%%%%%%%%%%%%%%%%%%%%%%%%%%%%%%%%%%%%%%%%%%%%%%%%%%%%%%%%%%%%%%%
\begin{figure}[t]
\begin{center}
\includegraphics[height=4.0cm]{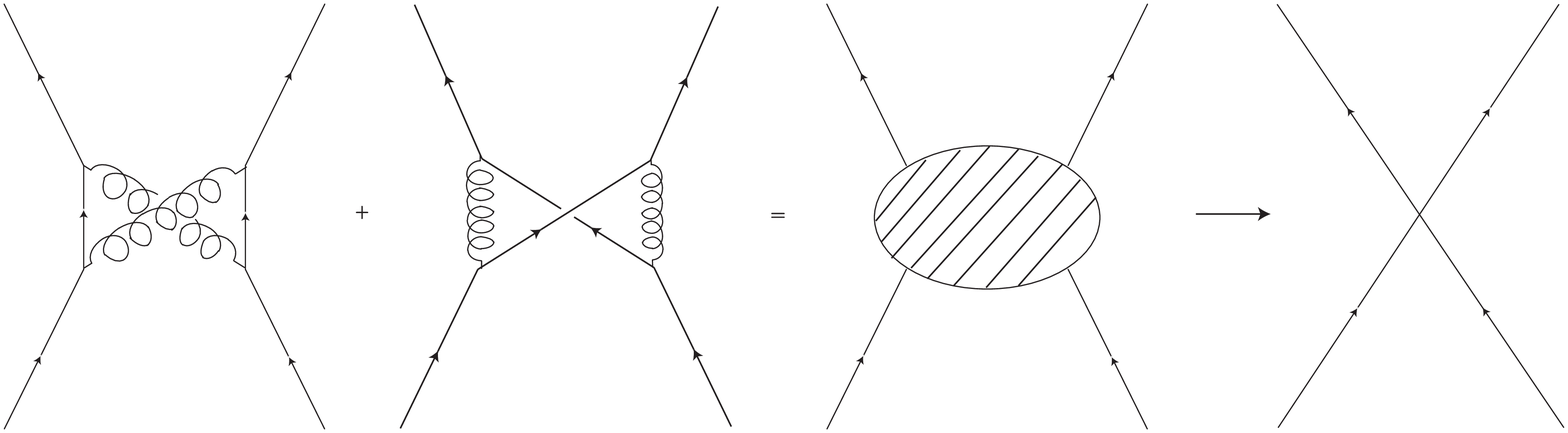}
\caption{
The Feynman diagrams of two-gluon exchange process. 
}
\label{fig:a2}
\end{center}
\end{figure}
%%%%%%%%%%%%%%%%%%%%%%%%%%%%%%%%%%%%%%%%%%%%%%%%%%%%%%%%%%%%%%%%%%%%%%%%
%

Finally, we obtain the effective Lagrangian density originated from the 
properly two-gluon exchange contribution between quarks, as is illustrated in Fig.\ref{fig:a2}, as follows:
\beq\label{d13}
\int d^4x {\cal L}_{\rm eff}
&=&g^4\int d^4x d^4y d^4x' d^4 y' ({\cal L}_{(c)}+{\cal L}_{(d)})\nonumber\\
&=&\int d^4x \left(
-\frac{G_T}{4}{\bar \psi}(x)\gamma^\mu \gamma^{\nu}\psi(x)\cdot {\bar \psi}(x)
\gamma_{\mu}\gamma_{\nu}\psi(x)+2G_T{\bar \psi}(x)\psi(x)\cdot {\bar \psi}(x)\psi(x)\right)\ , 
\nonumber\\
& &
\eeq
where we define $G_T\equiv 8g_T$ where $G_T>0$. 
We take $G_T=11$ GeV$^{-2}$ in this paper.
The first term corresponds to our tensor-type four-point interaction. 
Introducing the degree of freedom of the flavor, the tensor-part is written as 
\beq\label{d14}
{\cal L}_T=-\frac{G_T}{4}({\bar \psi}\gamma^\mu\gamma^\nu{\vect \tau}\psi)
({\bar \psi}\gamma_\mu\gamma_\nu {\vect \tau}\psi) \ .
\eeq
which is identical with the first term in Eq.(2.4). 
Of coure, the above treatment is nothing but a crude approximate treatment. 
Thus, we have to add another term so as to retain the chiral symmetry that the QCD has.  
The second term in Eq. (\ref{d13}) represents the scalar-scalar interaction appearing in the original 
NJL model Lagrangian. 
However, the one-gluon exchange contribution may be wash out this contribution from 
two-gluon exchange.

\vspace{-0cm}

%\appendix

%\section{}

% can use a bibliography generated by BibTeX as a .bbl file
% BibTeX documentation can be easily obtained at:
% http://www.ctan.org/tex-archive/biblio/bibtex/contrib/doc/

%\bibliographystyle{ptephy}
%\bibliography{sample}

\begin{thebibliography}{9}

\bibitem{FH}
See, for example, K. Fukushima and T. Hatsuda, Rep. Prog. Phys. {\bf 74}, 014001 (2011).

\bibitem{CS}
M. G. Alford, A. Schmitt, K. Rajagopal and T. Schafer, Rev. Mod. Phys. {\bf 80}, 1455 (2008) and references cited therein.



\bibitem{ARW}
M. Alford, K. Rajagopal and F. Wilczek, Nucl. Phys. B {\bf 537}, 443 (1999).

\bibitem{IB}
K. Iida and G. Baym, Phys. Rev. D {\bf 63}, 074018 (2001). 

\bibitem{MaC}
L. McLerran and R. D. Pisarski, Nucl. Phys. A {\bf 796}, 83 (2007). 


\bibitem{NT}
E. Nakano and T. Tatsumi, Phys. Rev. D {\bf 71}, 114006 (2005). 

\bibitem{Nic}
D. Nickel, Phys. Rev. Lett. {\bf 103}, 072301 (2009). 

\bibitem{Bub}
M. Buballa and S. Carignano, Prog. Part. Nucl. Phys. {\bf 81}, 39 (2015), and references cited therein.


\bibitem{oursPTEP1}
Y. Tsue, J. da Provid\^encia, C. Provid\^encis, M. Yamamura and H. Bhor, 
Prog. Theor. Exp. Phys. {\bf 2013}, 103D01 (2013).




\bibitem{oursPTEP2}
Y. Tsue, J. da Provid\^encia, C. Provid\^encis, M. Yamamura and H. Bhor, 
Prog. Theor. Exp. Phys. {\bf 2015}, 013D02 (2015).




\bibitem{oursPTEP3}
Y. Tsue, J. da Provid\^encia, C. Provid\^encis, M. Yamamura and H. Bhor, 
Prog. Theor. Exp. Phys. {\bf 2015}, 103D01 (2015).



\bibitem{magnetar1}
R. C. Duncan and C. Thompson, Astrophys. J. {\bf 392}, L9 (1992).

\bibitem{magnetar2}
C. Thompson and R. C. Duncan, Astrophys. J. {\bf 408}, 194 (1993).

\bibitem{magnetar3}
C. Thompson and R. C. Duncan, Astrophys. J. {\bf 473}, 322 (1996).

\bibitem{oursPTP}
Y. Tsue, J. da Provid\^encia, C. Provid\^encis and M. Yamamura, 
Prog. Theor. Phys. {\bf 128}, 507 (2012).







\bibitem{NJL}
Y. Nambu and G. Jona-Lasinio, Phys. Rev. {\bf 122}, 345 (1961), Phys. Rev {\bf 124}, 246 (1961).

\bibitem{HK}
T. Hatsuda and T. Kunihiro, Phys. Rep. {\bf 247}, 221 (1994).


\bibitem{Buballa}
M. Buballa, Phys. Rep. {\bf 407}, 205 (2005). 


\bibitem{arXiv}
H. Bohr, P. K. Panda, C. Provid\^encia and J. da Provid\^encia, 
Int. J. Mod. Phys. E {\bf 22}, 1350019 (2013).

\bibitem{NMT}
E. Nakano, T. Maruyama and T. Tatsumi, Phys. Rev. D {\bf 68}, 105001 (2003). 


\bibitem{TMN}
T. Tatsumi, T. Maruyama and E. Nakano, Prog. Theor. Phys. Suppl. No. 153, 190 (2004).



\bibitem{tensor}
M. Jaminon, E. Ruiz Arriola, Phys. Lett. B {\bf 443}, 33 (1998).

\bibitem{tensor2}
M. Chizhov, JETP Lett. {\bf 80}, 73 (2004). 


\bibitem{Ferrer}
E. J. Ferrer, V. de la Incera, I. Portillo and M. Quiroz, Phys. Rev. D {\bf 89}, 085034 (2014). 


%\bibitem{2}
%H. Bohr, C. Provid\^encia and J. da Provid\^encia, Eur. Phys. J. A {\bf 41}, 355 (2009).


%\bibitem{3}
%H. Bohr, P. K. Panda, C. Provid\^encia and J. da Provid\^encia, Braz.\ J. Phys. {\bf 42}, 68 (2012).






\end{thebibliography}
%
% once the .bbl file has been generated then place the text in your article.

%\vfill\pagebreak

\end{document}